\documentclass[twocolumn]{aastex63}

\newcommand{\w}{$W_r$~}
\newcommand{\kms}{km s$^{-1}$}
\newcommand{\z}{$z$}
\newcommand{\HI}{H~{\sc i}}
\newcommand{\MGII}{Mg~{\sc ii}}
\newcommand{\CIV}{C~{\sc iv}}

\received{}
\revised{}
\accepted{}
\submitjournal{ApJ}
 
\shorttitle{DESI-SV GQP}
\shortauthors{Zou et al.}

\begin{document}

\title{DESI survey validation data in the COSMOS/HSC field: Cool gas trace main sequence star-forming galaxies at the cosmic noon}

\correspondingauthor{Siwei Zou}
\email{siwei1905@gmail.com}

\author[0000-0002-3983-6484]{Siwei Zou}
\affiliation{Kavli Institute for Astronomy and Astrophysics, Peking University, Beijing 100871, China}
\affiliation{Department of Astronomy, Tsinghua University, Beijing 100084, China}

\author[0000-0003-4176-6486]{Linhua Jiang}
\affiliation{Kavli Institute for Astronomy and Astrophysics, Peking University, Beijing 100871, China}

\author[0000-0001-8467-6478]{Zheng Cai}
\affiliation{Department of Astronomy, Tsinghua University, Beijing 100084, China}

\author[0000-0002-2733-4559]{John Moustakas}
\affiliation{Department of Physics and Astronomy, Siena College, 515 Loudon Road, Loudonville, NY 12211, USA}

\author[0000-0002-8246-7792]{Zechang Sun}
\affiliation{Department of Astronomy, Tsinghua University, Beijing 100084, China}

\author[0000-0003-0230-6436]{Zhiwei Pan}
\affiliation{Kavli Institute for Astronomy and Astrophysics, Peking University, Beijing 100871, China}

\author[0000-0003-4651-8510]{Jiani Ding}
\affiliation{Department of Astronomy and Astrophysics, University of California, Santa Cruz, 1156 High Street, Santa Cruz, CA 95065, USA}
\affiliation{University of California Observatories, 1156 High Street, Sana Cruz, CA 95065, USA}

\author[0000-0002-2890-3725]{Jaime E Forero-Romero}
\affiliation{Departamento de F\'isica, Universidad de los Andes, Cra. 1 No. 18A-10, Edificio Ip, CP 111711, Bogot\'a, Colombia}

\author[0000-0002-6684-3997]{Hu Zou}
\affiliation{National Astronomical Observatories, Chinese Academy of Sciences, A20 Datun Rd., Chaoyang District, Beijing, 100012, P.R. China}

\author[0000-0001-5082-9536]{Yuan-sen Ting}
\affiliation{Research School of Astronomy \& Astrophysics, Australian National University, Cotter Road, Weston, ACT 2611, Australia}
\affiliation{School of Computing, Australian National University, Acton ACT 2601, Australia}

\author{Matthew Pieri}
\affiliation{Aix Marseille Univ, CNRS, CNES, LAM, Marseille, France}

\author[0000-0001-6098-7247]{Steven Ahlen}
\affiliation{Physics Dept., Boston University, 590 Commonwealth Avenue, Boston, MA 02215, USA}

\author[0000-0002-5896-6313]{David Alexander}
\affiliation{Centre for Extragalactic Astronomy, Department of Physics, Durham University, South Road, Durham, DH1 3LE, UK}

\author[0000-0002-8458-5047]{David Brooks}
\affiliation{Department of Physics \& Astronomy, University College London, Gower Street, London, WC1E 6BT, UK}

\author[0000-0002-4928-4003]{Arjun Dey}
\affiliation{NSF's NOIRLab, 950 N. Cherry Ave., Tucson, AZ 85719, USA}

\author[0000-0002-3033-7312]{Andreu Font-Ribera}
\affiliation{Institut de F\'{i}sica d’Altes Energies (IFAE), The Barcelona Institute of Science and Technology, Campus UAB, 08193 Bellaterra Barcelona, Spain}

\author[0000-0003-3142-233X]{Satya Gontcho A Gontcho}
\affiliation{Lawrence Berkeley National Laboratory, 1 Cyclotron Road, Berkeley, CA 94720, USA}

\author[0000-0002-6550-2023]{Klaus Honscheid}
\affiliation{Department of Physics, The Ohio State University, 191 West Woodruff Avenue, Columbus, OH 43210, USA}
\affiliation{Center for Cosmology and AstroParticle Physics, The Ohio State University, 191 West Woodruff Avenue, Columbus, OH 43210, USA}

\author[0000-0003-1838-8528]{Martin, Landriau}
\affiliation{Lawrence Berkeley National Laboratory, 1 Cyclotron Road, Berkeley, CA 94720, USA}

\author{Axel de la Macorra}
\affiliation{Instituto de F\'{\i}sica, Universidad Nacional Aut\'{o}noma de M\'{e}xico, Cd. de M\'{e}xico C.P. 04510, M\'{e}xico}

\author[0000-0003-3841-1836]{Mariana Vargas Magana}
\affiliation{Instituto de F\'{\i}sica, Universidad Nacional Aut\'{o}noma de M\'{e}xico, Cd. de M\'{e}xico C.P. 04510, M\'{e}xico}

\author[0000-0002-1125-7384]{Aaron Meisner}
\affiliation{NSF's NOIRLab, 950 N. Cherry Ave., Tucson, AZ 85719, USA}

\author{Ramon Miquel}
\affiliation{Institut de F\'{i}sica d’Altes Energies (IFAE), The Barcelona Institute of Science and Technology, Campus UAB, 08193 Bellaterra Barcelona, Spain}
\affiliation{Instituci\'{o} Catalana de Recerca i Estudis Avan\c{c}ats, Passeig de Llu\'{\i}s Companys, 23, 08010 Barcelona, Spain}

\author[0000-0001-9504-2059]{Michael Schubnell}
\affiliation{Department of Physics, University of Michigan, Ann Arbor, MI 48109, USA}

\author[0000-0003-1704-0781]{Gregory Tarl\'e}
\affiliation{University of Michigan, Ann Arbor, MI 48109, USA}

\author[0000-0002-4135-0977]{Zhimin Zhou}
\affiliation{National Astronomical Observatories, Chinese Academy of Sciences, A20 Datun Rd., Chaoyang District, Beijing, 100012, P.R. China}

%\author{other co-authors}

%%%%%%%%
\begin{abstract} 
We present the first result in exploring the gaseous halo and galaxy correlation using the Dark Energy Spectroscopic Instrument (DESI) survey validation data in the Cosmic Evolution Survey (COSMOS) and Hyper Suprime-Cam (HSC) field. We obtain the multiphase gaseous halo properties in the circumgalactic medium (CGM) by using 115 quasar spectra (S/N $>$ 3). We detect \MGII~absorption at redshift 0.6 $< z <$ 2.5, \CIV~absorption at 1.6 $<z<$ 3.6, and H~{\sc i} absorption associated with the \MGII~and \CIV. By cross-matching the COSMOS2020 catalog, we identify the \MGII~ and \CIV~host galaxies in ten quasar fields at 0.9 $< z <$ 3.1. We find that within the impact parameter of 250 kpc, a tight correlation is seen between strong \MGII~equivalent width and the host galaxy star formation rate. The covering fraction $f_c$ of strong \MGII~selected galaxies, which is the ratio of absorbing galaxy in a certain galaxy population, shows significant evolution in the main-sequence galaxies and marginal evolution in all the galaxy populations within 250 kpc at $0.9 < z<$ 2.2. The $f_c$ increase in the main-sequence galaxies likely suggests the co-evolution of strong \MGII~absorbing gas and the main-sequence galaxies at the cosmic noon. Furthermore, several \MGII~and \CIV~absorbing gas is detected out of the galaxy virial radius, tentatively indicating the feedback produced by the star formation and/or the environmental effects.

\end{abstract}


%%%%%%%%%%%%%%%%%%%
eh \section{Introduction}\label{sec_intro}
The baryon cycle in and around a galaxy is of critical importance in understanding the cosmic star formation history and galaxy evolution. The so-called interstellar medium (ISM) and circumgalactic medium (CGM) play a key role in regulating the baryon cycle and thus galaxy evolution (see reviews of \citealt{tum17,per20} and the references therein). The absorption lines produced by the intervening medium toward bright background quasars provide a sensitive measurement of the multi-phase gas properties in the transverse direction of the gaseous halos. Different phases (density, temperature, ionization parameter) of the gas in the ISM and CGM can be characterized by different ions. For instance, neutral atomic carbon (C~{\sc i}) is used to trace cold, metal-enriched \citep{led15,zou18} and molecular gas \citep{not18}. The Ly$\alpha$ absorption systems can probe optically thick neutral gas, known as the damped Lyman$\alpha$ systems (neutral hydrogen column density N (H~{\sc i}) $>10^{20.3}$ cm$^{-2}$, DLA) (\citealt{pet00a,wol05,pro06,not11,raf12,kro17,nee19,lin22}). Mg~{\sc ii} ($\lambda\lambda$2796,2803) and \CIV~doublets are found to reside in the cool (T$\sim$ 10$^4$ K) (\citealt{ber91,ste02}) and warm-hot gas (T$\sim$ 10$^{5.5-6}$ K) gas, respectively \citep{bor14,bur16}. Additionally, peculiar abundance patterns of DLA or Lyman-Limit-Systems potentially exhibit the signature of the old generation of stars (e.g., \citealt{zou20,welsh22}).

Extensive experiments have been designed to explore the CGM-galaxy correlation and its role in galaxy evolution at redshift $z<$ 1. The CGM has been found to co-rotate with the host galaxy along the major and minor axis \citep{nie13a}. The strength (equivalent width or column density) of the cool gas tentatively correlates with the luminosity ($L_B$ of the host galaxy \citep{chen10} and is anti-correlated with the impact parameter of the galaxy \citep{bou06,nie13b}. %The DLA selected galaxy have only been found $\sim$ 20 in the past 20 years \citep{kro17}.
Emission from the diffuse gas region has also been detected in recent work (e.g., \citealt{fel18,lec22}).

Observations are still limited in exploring the CGM-galaxy correlation at $z>$ 1 because direct galaxy observation is very challenging for ground-based telescopes. Stacking of a large quantity of Sloan Digital Sky Survey data (SDSS) quasar spectra can be used to probe the weak galaxy emission at high redshift \citep{jos17} and CGM distribution at a large scale \citep{pie14}. Recently, integral field spectroscopy such as Keck Cosmic Web Imager (KCWI) and VLT/Multi Unit Spectroscopic Explorer (MUSE) provides an efficient tool to provide a 3D view in both the large and small-scale of the CGM at $z < 1.5$ and $z>2$. Samples of Mg~{\sc ii}-galaxy pairs at $z< 1.5$ are built using VLT/MUSE \citep{zab19,sch19,sch21,dutta20}. A bimodality of azimuthal angle-metallicity relation has been found in MusE GAs FLOw and Wind (MEGAFLOW) survey and at $z\sim$1 to trace either inflow \citep{zab19} or galactic outflow \citep{sch19}. The environmental effects of cool gas traced by \MGII~have been reported in the MUSE Analysis of Gas around Galaxies (MAGG) survey \citep{dutta20}.

At $z$ = 2--4, the CGM can be probed by the emission lines near bright quasars (e.g., \citealt{can14,mar14a,cai19,fos21}) and overdense regions \citep{cai17a,cai17b}. Surveys to study the small-scale CGM structure and its correlation with the host galax(ies) are somehow scarce (e.g., the Keck Baryonic Structure Survey, \citealt{rud12,rud19}). The cosmic star formation rate and baryon accretion peak around $z\sim$ 2 \citep{madau14}. In order to trace the multiphase gas-galaxy co-evolution at $z$ = 1--3, especially towards the cosmic noon ($z = $ 2--3), we present a pilot study that takes advantage of the 30 multiband photometry in the Cosmological Evolution Survey (COSMOS) field to search for the CGM host galaxies. In \citet{zou21} (hereafter Z21), the authors tentatively search for strong Mg~{\sc ii} absorbers (2 $<z <$ 6) counterparts from Hubble Space Telescope/Canada France Hawaii Telescope/The Dark Energy Camera Legacy Survey deep images. The result indicates that the strong Mg~{\sc ii} absorbing gas tends to have a smaller halo size but a more disturbed environment than that at lower redshift. 

In this paper, we will first present CGM gas properties by using the quasars observed in the  Dark Energy Spectroscopic Instrument (DESI) survey validation (SV) in the COSMOS and Hyper Suprime-Cam Subaru Strategic Program (HSC) fields, then we will particularly study the CGM-galaxy correlation in the COSMOS field. This paper is presented as follows: we introduce the observation and data analysis in Section \ref{sec_obs}. The multiphase gas properties are presented in Section \ref{sec:multiphase}, and the gas-galaxy correlation is presented in Section \ref{sec:gas_galaxy}. We discuss and summarize the implication of this work in Section \ref{sec:discussion} and \ref{sec:summary}. 

\section{Observation and Data Processing}\label{sec_obs}

DESI is a Stage IV ground-based dark energy experiment that studies baryon acoustic oscillations and the growth of structure through redshift-space distortions with a wide-area galaxy and quasar redshift survey \citep{desia,desib,desi22}. Full details of the DESI early data release and secondary projects are described in \citet{edr_a,edr_b}. Descriptions of the SV data and data reduction pipeline are presented in \citet{desi23_sv} and \citet{desi_pipeline}, respectively. Target selection and samples of quasars, bright galaxies, emission-line galaxies, and luminous red galaxies from the SV data can be found in \citet{yeche20,chau22,rai22,yang23,zhou23,lan23}. Full \MGII~absorber catalog of the DESI EDR data is presented in \citet{desi_mg_nap}.

This project is a DESI secondary program in the SV. The quasars are selected from the overlapping region of COSMOS and HSC Ultra-Deep \citep{aih18} fields (2 deg$^2$ and having the tangent point RA, DEC at 150.116, 2.210). Observations of quasars in this work were conducted at Kitt Peak by DESI between March 2021 to May 2021. The average effective exposure time 
%observation hour 
is around 5.5 hours. The DESI SV quasar catalog is created with three algorithms: the DESI pipeline classifier Redrock (RR, Bailey et al. in prep), a broad Mg~{\sc ii} line finder, and a machine learning-based classifier QuasarNET \citep{busca18,farr20}. The RR algorithm is a template-fitting classifier to classify the quasars using the templates of the different targets (stars, galaxies, and quasars) generated from the SDSS spectra. The Mg~{\sc ii} line finder is an afterburner algorithm using the inputs and outputs from RR. QuasarNET is a deep convolutional neural network classifier. Details of the quasar visual inspection result are presented in \cite{ale22}.

\begin{figure}
  \resizebox{\hsize}{!}{\includegraphics{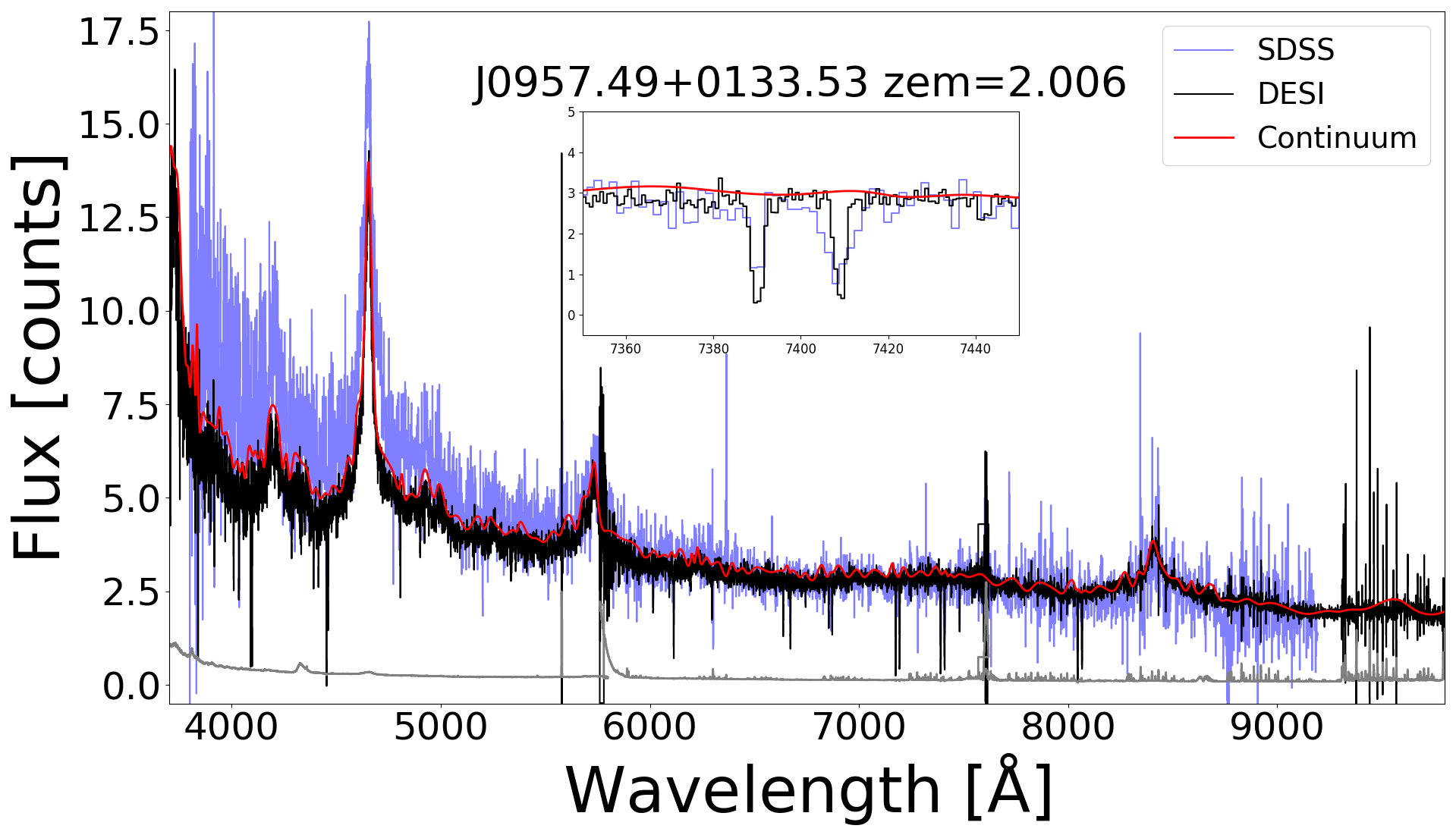}}
  \caption{\small{Spectrum of J095749.98+013354.1 taken by DESI (black) and SDSS (blue), respectively. Zoom in spectra of one Mg~{\sc ii} detected at $z$ = 1.6428. The exposure time of DESI and SDSS observation is 7302s and 4027s, respectively. The red curve is the continuum fitting done by the method described in Section \ref{sec:norm}.
}}\label{fig:desi_sdss}
\end{figure}

From all the observed quasars in the COSMOS/HSC field, we selected 115 quasars each with a mean signal-to-noise ratio (S/N) greater than 3. We calculate the mean S/N of one spectrum as the mean S/N per pixel from three different intervals, where there are no significant emission and absorption lines (flux residuals smaller than 0.5). Details of all quasars used in this work are presented in Appendix Table \ref{table:qso_info}. Among these 114 quasars, 30 quasars are included 
in the quasar catalog of SDSS release 16 \citep{lyke20} and the rest 84 are new quasars. We show the spectra comparison of one quasar (J095749.98+013354.1) taken by DESI and SDSS is presented in Figure \ref{fig:desi_sdss}. The DESI spectra resolution at 3600 -- 9800 \AA~ranges from 2000 to 5000.  
 
\subsection{Normalization and metal lines finder}\label{sec:norm}

To detect the intervening absorption systems, we first normalize the quasar spectrum by dividing the flux with the continuum. The continuum defined in this work is the spline line without a significant absorption feature (see the red curve line in Figure \ref{fig:desi_sdss} for an example). We use two methods to estimate the quasar continuum: the principle component analysis (PCA) method (see e.g., \citealt{paris11,guo18}) and an unsupervised probabilistic continuum fitting method with uncertainty quantification \citep{sun22}\footnote{https://github.com/ZechangSun/QFA/}. The fitting spectra redward Ly$\alpha$ emission from these two methods are consistent. The fitting result is then visually checked to avoid any significant improper fitting. 

After obtaining the normalized spectra, we use the algorithm described in Z21 to automatically search for metal lines redward of Ly$\alpha$ emission, i.e., rest frame $\lambda$ 1216 -- 3000 \AA. We briefly introduce the algorithm here. We identify the absorption feature from the normalized spectrum with a Gaussian kernel filter having rest-frame velocity FWHM smaller than 600 km s$^{-1}$. When the rest-frame equivalent width $W_r$ of this Gaussian kernel is greater than our detection limit (0.3 \AA), this kernel is then identified as an absorption feature. We search for the Mg~{\sc ii} (C~{\sc iv}) doublets with two Gaussian kernels that are separated around 770 km s$^{-1}$ (500 \kms). The self-blending of \MGII~and \CIV~systems can be mitigated by the two Gaussian kernels. We then visually inspect all the detected absorbers to ensure there is no significant blending from absorber systems at other redshifts. The error of the equivalent width is calculated by the flux variance summation over the search boxcar. We also add the SNR constraint in the nearby region of the absorption feature. The final absorber sample is visually checked. The criteria to select the \MGII~(C~{\sc iv}) doublets are as follows:

\noindent(1) the local S/N $>3$; where the local S/N is the mean S/N per pixel around $\pm$ 10 pixels adjacent to the search boxcar center. \\
(2) $W_r (\lambda2796)> $ 0.3 \AA~and $W_r (\lambda2803)>$ 0.15 \AA~($W_r (\lambda1548)> $ 0.3 \AA~and $W_r (\lambda1550)>$ 0.15 \AA) for C~{\sc iv}; \\
(3) $W_r(\lambda2796)$ / $\sigma$ ($W_r$) $>$ 3 and $W_r(\lambda2803)$ / $\sigma$ ($W_r$) $>$ 3 ($W_r(\lambda1548)$ / $\sigma$ ($W_r$) $>$ 3 and $W_r(\lambda1550)$ / $\sigma$ ($W_r$) $>$ 3 for C~{\sc iv}) indicating a 3$\sigma$ detection.
%{\bf please read again to confirm this sentence.}. 

\section{Multi-phase gas}\label{sec:multiphase}

The multiplicity of the absorbers is detected (\HI, \MGII, \CIV, Al~{\sc ii}, Al~{\sc iii}, Si~{\sc ii}, Si~{\sc iii} and Si~{\sc iv}) from the DESI SV quasar spectra. In this work, we %will 
focus on the discussion of \HI, \MGII~and \CIV. We detect \MGII~and \CIV~independently using the algorithm described in Section \ref{sec:norm}, the \HI~systems are then searched in the same system once the \MGII~or \CIV~are detected. 

Quantitatively, we detect {\bf 51} \MGII~absorption (0.66 $<z<$ 2.49, 0.27 $<W_r<$ 4.89 \AA) and {\bf 50} C~{\sc iv} absorption (1.35 $<z<$ 3.18, 0.22 $< W_r < $ 3.05 \AA).  In the overlapping redshift coverage between Mg~{\sc ii} and H~{\sc i} ($z =$ 1.95 -- 2.50); \CIV~and \HI~($z =$ 1.95 -- 3.20), %there are 
8 out of 10 (80\%) Mg~{\sc ii} systems have \HI~detection, and 24 out of 29 (82.75\%) \CIV~systems have \HI~detection. 
In the redshift coverage of both \MGII~and \CIV~($z =$ 1.3 -- 2.5), 20 out of 34 (58.82\%) Mg~{\sc ii} systems have C~{\sc iv} detection. We found a relatively larger median $W_r$ of the \CIV~systems with \MGII~absorption (0.859 \AA) than \CIV~systems without \MGII~absorption (0.694 \AA). 

\begin{figure}
  \resizebox{\hsize}{!}{\includegraphics{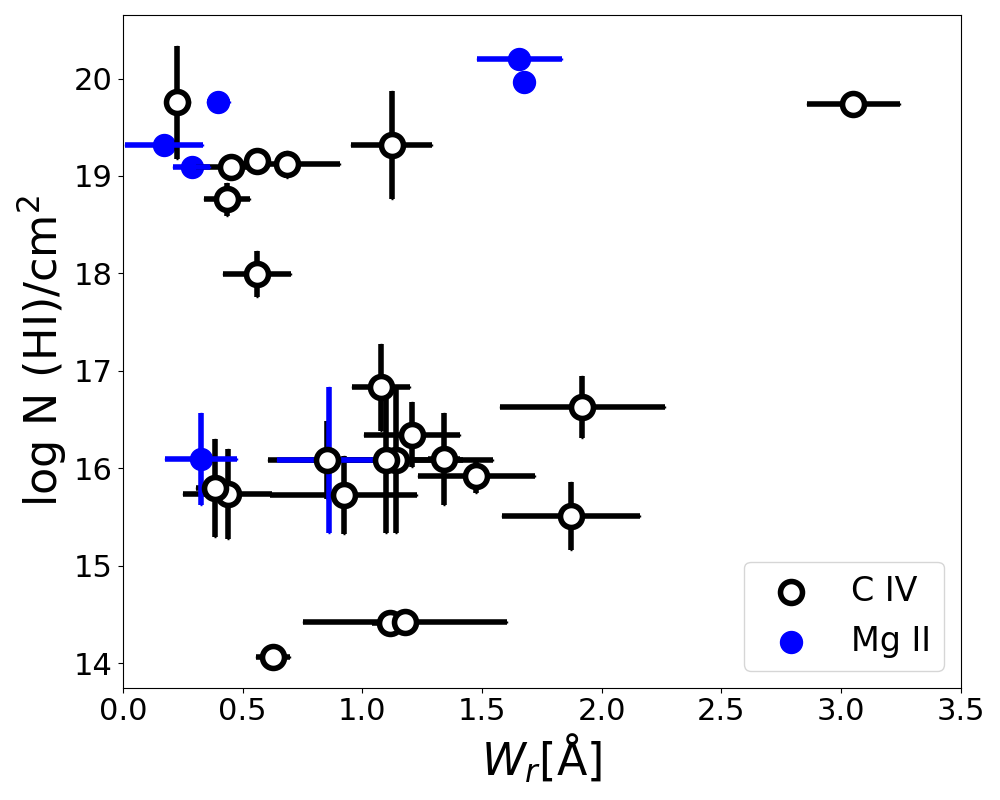}}
  \resizebox{\hsize}{!}{\includegraphics{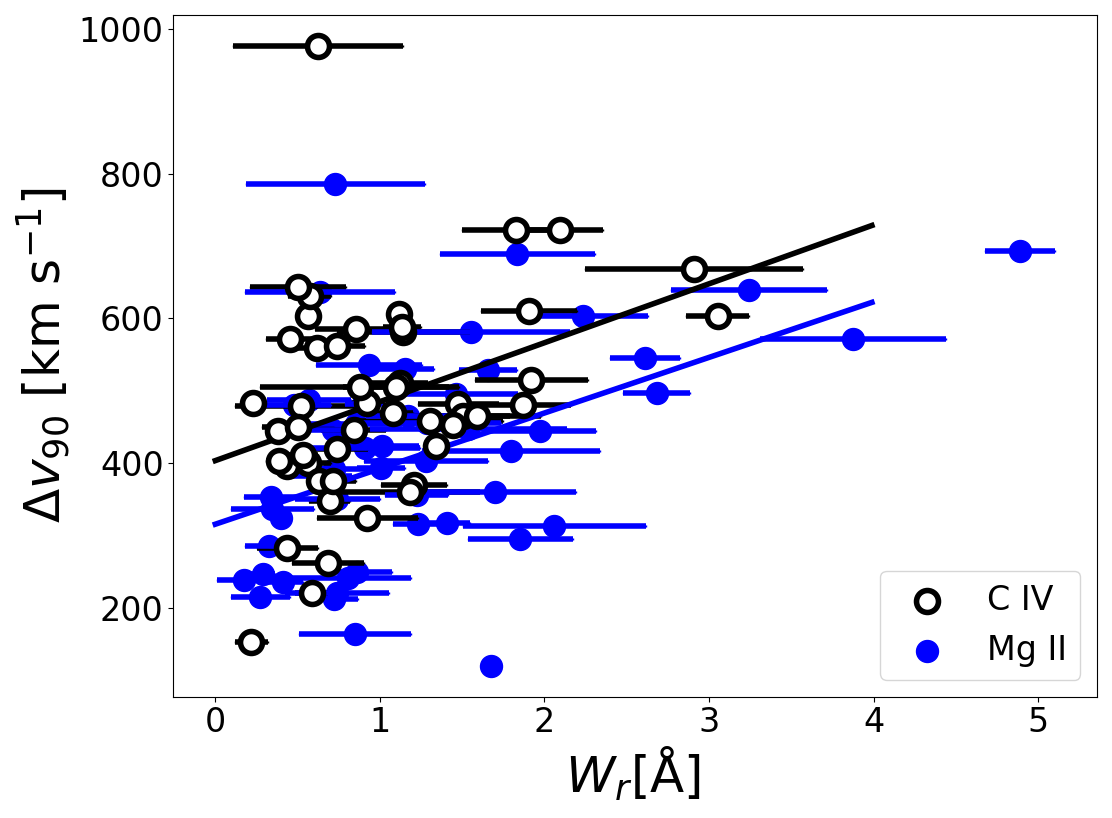}}
  \caption{\small{$Upper$ The column density of H~{\sc i} against the equivalent 
  widths of \MGII (blue dots) and \CIV~(hollow black circles) where the three ions 
  are detectable. All of the \CIV~and \MGII~systems have log $N$(H~{\sc i})/cm$^{2}$
  $> 14.0$. The \MGII~systems have relatively higher $N$(H~{\sc i}) than \CIV~systems. $Lower$ The $\Delta v_{90}$-$W_r$ relation of our detected \MGII~and \CIV~absorbers. The $\Delta v_{90}$ is the velocity spread encompassing 90\% of the total optical depth of the absorption line. The Pearson coefficient $p$ value for the $\Delta v$ of \MGII~and \CIV~absorbers is 0.024, indicating a similarity between these two variables.
}}\label{fig:CIV_HI}
\end{figure}

We plot all systems having \HI~detection in the upper panel of Figure \ref{fig:CIV_HI}. 
The figure indicates that all of the detected H~{\sc i} absorption 
associated with \MGII~and \CIV~systems have 
log $N$(H~{\sc i})/cm$^{2}$ $>$ 14.0. The column density of H~{\sc i} is measured using the Voigt profile (package Voigtfit, \citealt{kro18}) when there are strong Lyman series lines (e.g., $W_r(\lambda1215) >$ 10 \AA) detected at the redshifts of metal line absorber. The total column density of H~{\sc i} is derived from both the Ly$\alpha$ and Ly$\beta$ lines (when detected). Both Ly$\alpha$ and Ly$\beta$ lines are fitted with the same Doppler parameter $b$, which is varied between 40-200 km s$^{-1}$. The final $N$ value and its associated error are taken when the smallest $\chi^2$ value is obtained. The complete fit results of \MGII, \CIV~and H~{\sc i} systems 
are presented in Appendix Table \ref{table:metal_measurement}. It suggests that \MGII~absorbers probe a relatively higher $N$(H~{\sc i}) gas than the \CIV~absorbers. All of the \MGII~systems have log $N$(H~{\sc i})/cm$^{2}$ $>$ 16.0, which is consistent with what previous detections have claimed for \MGII~absorption at $z$ $<$ 1 (e.g., \citealt{chur00a,nie13a}). Particularly, five out of seven (71.43\%) \MGII~systems are associated with DLA and sub-DLA (19.0 $<$ log $N$(H~{\sc i})/cm$^{2}$ $<$ 20.3) systems, indicating a significant correlation between \MGII~systems and large $N$(H~{\sc i}) gas. The \CIV~systems are likely to probe wider range of $N$(H~{\sc i}) gas than \MGII: one is associated with a more neutral phase having 18.5 $<$ log $N$(H~{\sc i})/cm$^{2}$ $<$ 20 and one is associated with a lower density phase (14.0 $<$ log $N$(H~{\sc i})/cm$^{2}$ $<$ 17.0). Note that the system with the highest $N$(\HI)~has \MGII~but no 3$\sigma$ \CIV~detection.

\subsection{Line density evolution of different ions}
%\subsection{Survey completeness}\label{sec:completeness}

\begin{figure}
\gridline{\fig{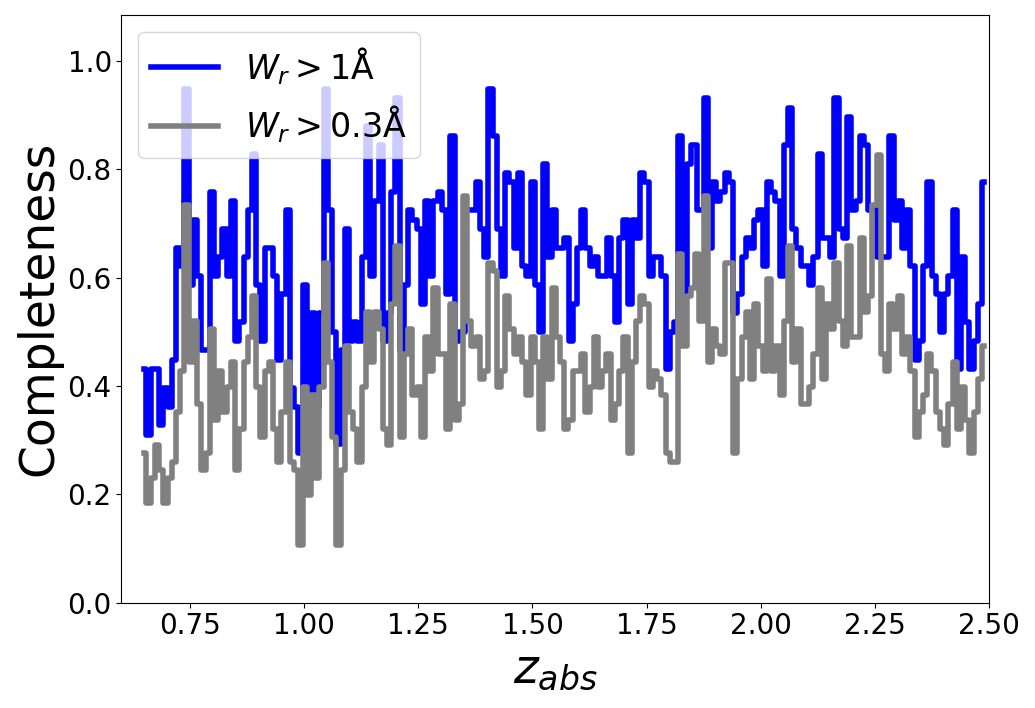}{0.5\textwidth}{(a)}
          }
\gridline{\fig{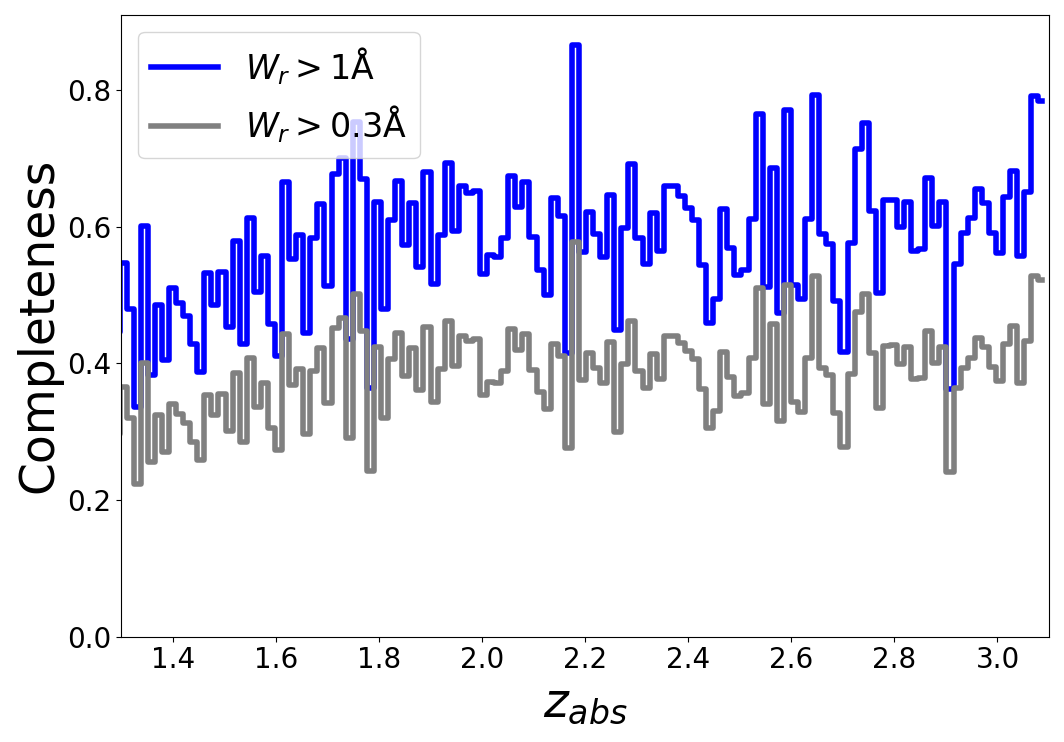}{0.5\textwidth}{(b)} 
          }
\caption{Average pathlength-weighted completeness as a function of absorbing redshift ($z_{abs}$) of \MGII~(upper) and \CIV~(lower) for 114 sightlines. The blue and grey curves are the completeness distribution with the equivalent width limits of $W_r \geq$ 1.0 \AA~and $W_r \geq$ 0.3 \AA, respectively.
\label{fig:completeness}}
\end{figure}

In this paragraph, we present the statistical properties of the metal absorbers. Firstly, 
we calculate the survey completeness by taking into account %of 
the contamination such as the skylines and spikes in the spectra. 
To calculate the survey completeness, we follow and update the algorithm described in Z21. 
We briefly introduce the algorithm here. The method is to sample uniformly 
mock Mg~{\sc ii} and C~{\sc iv} doublets in each spectrum using a Monte Carlo simulation. Then, we use the detection algorithm in Section \ref{sec:norm} to calculate the recovery rate. 
We generate 10,000 doublets with equivalent widths that vary uniformly distributed between 0.3 and 4.0 \AA. We measure the real velocity spread of the absorption profile, $\Delta v_{90}$, defined as the velocity spread containing 90\% of the total optical depth of the absorption line (e.g., \citealt{pro08}). We fit a relation between $W_r$ and velocity spread $\Delta v_{90}$ with a polynomial curve fitting technique considering the errors from two variables of \MGII~and \CIV~absorbers: $\Delta v_{90}$ = 77 $\times W_r(\lambda 2796)$ + 315 km s$^{-1}$ and $\Delta v_{90}$ = 81 $\times W_r(\lambda 1548)$ + 403 km s$^{-1}$. The $\Delta v_{90}$--$W_r$ relation of \MGII~and \CIV~is plotted in the lower pane of Figure \ref{fig:CIV_HI}. %The median velocity spread of \MGII~and \CIV~are 448 km/s and 535 km/s, respectively. 
The inserted observed velocity of the absorber is given by the $\Delta v_{90}$--$W_r$ relation. We then insert the mock doublets in 10,000 uniformly distributed redshifts in our searching path length (0.5 to 2.6 for Mg~{\sc ii} and 1.3 to 3.2 for C~{\sc iv}). The detection result of every inserted absorber is described as a Heaviside function $H(z,W_r)$. The path density $g(z,W_r)$ is the integral of $H(z,W_r)$ over the total sightline number $N$. The average completeness is then calculated by the average of path density $C = \int_{W_{lim}}^{\infty}g(z,W_r)/N$ as a function of redshift and $W_r$ limit.
(see Figure \ref{fig:completeness}).

\begin{figure}
  \resizebox{\hsize}{!}{\includegraphics{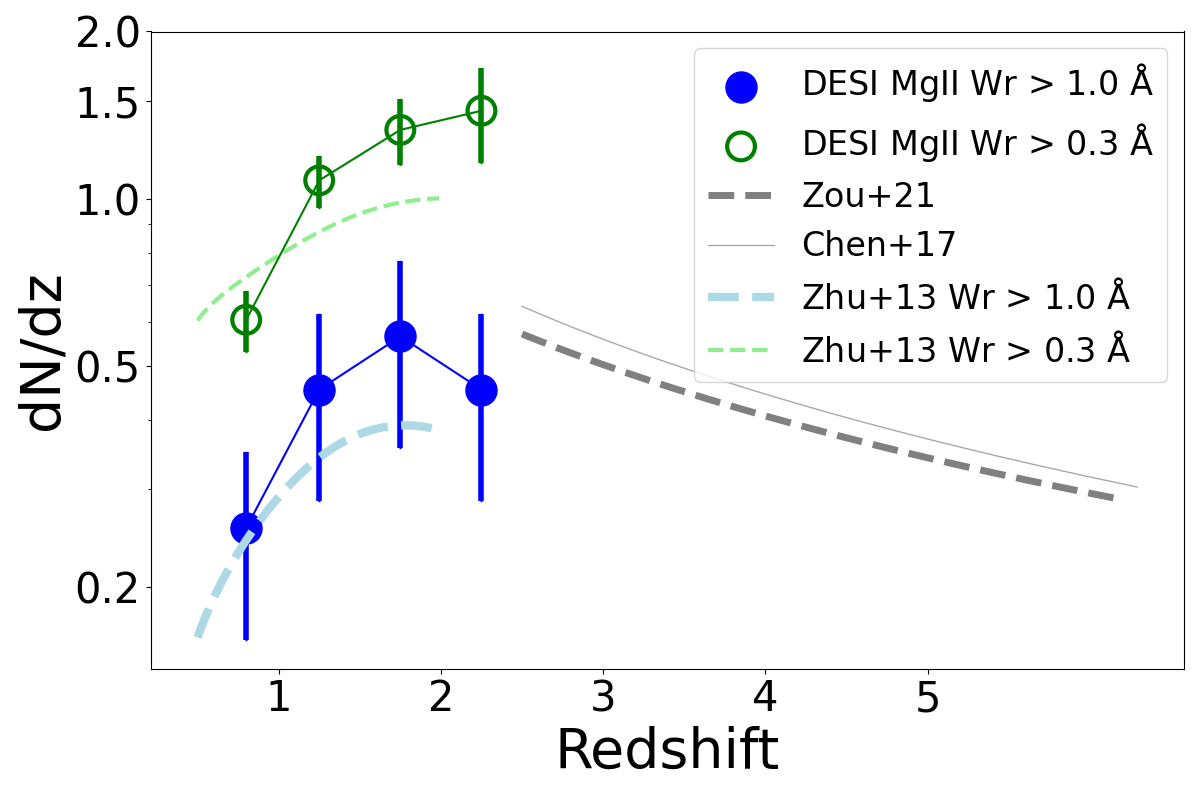}}
  \caption{\small{Line density $dN/dz$ of the \MGII~absorbers above 0.3 \AA\ (green circles) and 1.0 \AA\ (blue points) in this work. The grey dashed and solid lines are the fitting results of strong \MGII~($W_r > 1.0$ \AA) at $2.2 < z < 6.0$ in Z21 and \citet{chen17}, respectively. The strong \MGII~line density evolution is consistent with the cosmic star formation evolution from $1.0 <z< 6.0$. 
}}\label{fig:dndz_mg_civ}
\end{figure}

After applying the survey completeness correction, we calculate the line density per redshift bin ($dN/dz$) of \MGII~and \CIV. The line density of \MGII~and \CIV~follow the Poisson distribution \citep{zhu13a}, thus, the error of $dN/dz$ is given by 1/$\sqrt{N}$. The final line density of \MGII~are %is 
presented in Table \ref{table:dndz} and Figure \ref{fig:dndz_mg_civ}. We compare our calculated $dN/dz$ of \MGII~with {other} surveys at $z =$ 1--2 \citep{zhu13a} and $z>2$ \citep{chen17,zou21}. 
Figure \ref{fig:dndz_mg_civ} clearly shows that the line density evolution of strong \MGII~($W_r>$ 1 \AA) is consistent with the cosmic star formation history from the local Universe to $z\sim$ 6. Particularly, the trend has a tentative turnover at $z\sim$ 2. The line densities of weaker \MGII~systems (0.3 \AA $<W_r<$ 0.6 \AA, 0.6 \AA $<W_r<$ 1.0 \AA) increase with the increasing redshift from $z = 0.6-2.5$. 

\section{Absorbing gas-host galaxy correlation}\label{sec:gas_galaxy}

In this section, we first discuss the method of identifying the absorbing gas host galaxy. We %will 
then discuss the correlation between the gas phase properties (absorber strengths and ionization states) and the host galaxy properties (stellar mass, halo mass, Virial radius, impact parameter, and star formation rate (SFR)). {\bf Fourteen} quasar fields among the whole sample (115 quasars) are covered by the COSMOS2020 catalog \citep{weaver22}. We %will 
focus on the gas-galaxy correlation in the following discussion of these fourteen quasar fields. We %will also 
compare our results with other surveys such as the MAGIICAT \citep{nie13a,chur13a,chur13b} and the MEGAFLOW \citep{zab19,sch19,sch21} surveys (for \MGII~) at $z<1$ and $z =  1-1.5$, and the Cosmic Origin Spectrograph(COS) survey (for \CIV~) at $z<$ 1 for low-mass galaxies \citep{bort13,bor14}.

\subsection{Galaxy Identification}\label{sec:galaxy_identification}

The identification of the CGM gas host galaxies is one of the key processes in revealing the correlation between the multiphase gas and the host galax(ies). To compare our result with the MUSE surveys, we search for galaxy candidates within a radius of 30$\arcsec$ (257.51 kpc at $z = 2.0$) using the COSMOS2020 galaxy catalog \citep{weaver22}. 
COSMOS is the survey that contains half a million galaxies with a limiting magnitude of 27.2 AB in the F814W band. The radius of 30$\arcsec$ is determined to match the MUSE field of view of 1 arcmin. The sky coverage of our survey is 2 deg$^2$. 
Galaxies are detected in the combined $zYJHK_s$ images from the Subaru and Visible Infrared Survey Telescope for Astronomy (VISTA) telescopes (\citealt{sco07, lai16, weaver22}). 
By adding particularly the ultra-VISTA DR4 data and HSC PDR2 \citep{aih18}, the photometric redshift ($z_{phot}$) uncertainties in COSMOS2020 can match those of galaxies 0.7 magnitudes brighter in COSMOS2015 \citep{lai16}. The photometric redshift is measured by the LePhare \citep{arn02,ilb06} and EAZY \citep{bra08}. The likelihood distribution of the $z_{phot}$ is given by the  observed photometry and redshift $\mathcal{L}$(data$|z$). The final $z_{phot}$ is the median of the likelihood distribution ($z_{pdf}$) and the error bar is the 1$\sigma$ interval. Details of the photo-$z$ measurement are presented in \citet{weaver22}. The galaxy stellar mass and SFR are computed using LePhare with the same configuration as COSMOS2015 (see \citet{lai16}). Template library generated from the stellar population synthesis models in \citet{bc03} are fitted to the observed photometry to obtain the galaxy properties. We adopt the `best' value in the catalog which are taken at the minimum chi2. The stellar mass and SFR uncertainties are within the 68\% confidence level. In this work, we use THE FARMER v2.1 catalog\footnote {cosmos2020.calet.org}.

\begin{figure}
  \resizebox{\hsize}{!}{\includegraphics{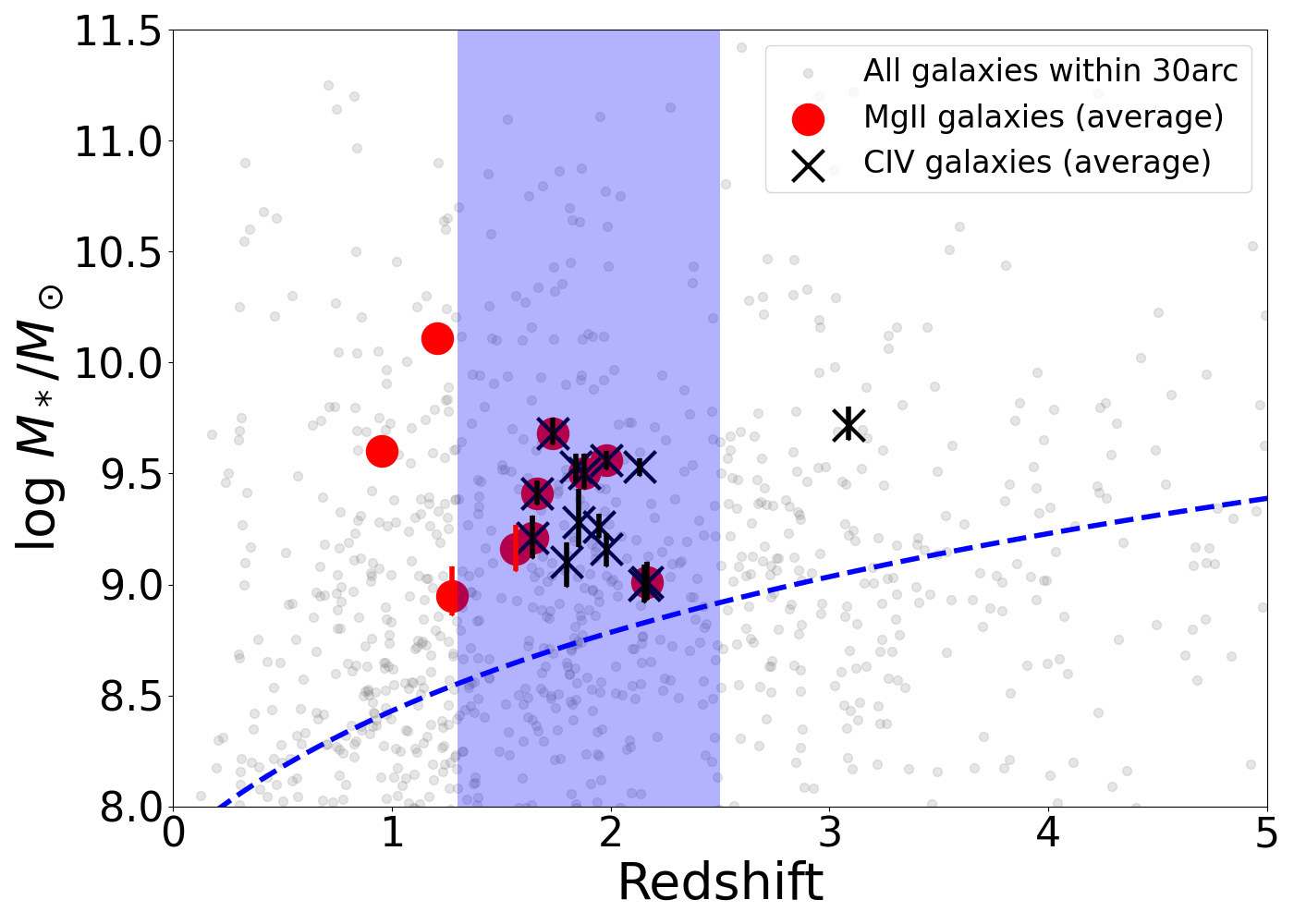}}
  \caption{\small{The stellar mass distribution of  
  galaxies in the 14 sightlines covered by the COSMOS2020 catalog. The black curve is the stellar mass limit as a function of redshift from \citet{weaver22}. %The 
  Red dots are \MGII~galaxy candidates, 
  and blue crosses are the \CIV~galaxy candidates. %The 
  Grey dots are all %the 
  galaxies within 30 $\times$ 30 arcseconds offset the quasar. The blue rectangle is the detection redshift coverage of both \MGII~and \CIV.
}}\label{fig:masslimit_mgii_civ}
\end{figure}

The precision of $z_{phot}$ is quantified with the difference between spectroscopic redshift $z_{spec}$ and photometric redshift: $\sigma = |z_{phot} - z_{spec}|/(1+z_{spec})$. We measure the $z_{phot}$ precision in the COSMOS field with the galaxy $z_{spec}$ from DESI SV data. We find that the median $\sigma$ is around 0.006. We select the absorbing gas host galaxies using criteria as follows: {\bf (a) } $\sigma <$ 0.01; 
{\bf (b)} we search the galaxy candidates within a $\pm$1000 km s$^{-1}$ velocity window with the absorber at $z_{abs}$ in the center, i.e., $c\times |z_{phot}$-$z_{abs}|$/(1+$z_{abs}) <$ 1000 km s$^{-1}$. %{\bf what is the meaning of the velocity window offset? }
Considering the photometric redshift uncertainty is still large for this velocity window ($\pm$1000 km s$^{-1}$), we include $all$ the galaxies whose $z_{pdf}$ distribution across this velocity window around the absorber's redshift instead of finding $the$ host galaxy. After applying these criteria, we pre-selected the Mg~{\sc ii} and C~{\sc iv} host galaxy candidates. %In the overlapping redshift 
%range of \CIV~and \MGII~ (blue rectangle in Figure~\ref{fig:masslimit_mgii_civ}), 
%five out of seven \CIV~galaxies have \MGII~absorption as well, 
%two \CIV-only galaxies have the lower \CIV~$W_r$ than %the 
%galaxies having both \MGII~and \CIV~. 
We plot all the pre-selected galaxy candidates in Figure \ref{fig:masslimit_mgii_civ}. The blue dashed line in Figure \ref{fig:masslimit_mgii_civ} is the stellar mass completeness calculated in \citet{weaver22}. We also test different galaxy selection criteria by cutting the stellar mass limit (blue dashed line in Figure \ref{fig:masslimit_mgii_civ}), SFR (log SFR $>$ --1) and $L_v$ ($L_v/L_* > $ 0.01, $L_v$ is the V-band luminosity and $L_*$ is the characteristic galaxy luminosity at $z\sim$ 2) limits, and find that the major result discussed in the following does not change.

\subsection{Galaxy counterparts properties}
We first present the \MGII~and \CIV~host galaxies properties in this section. In Section \ref{sec:galaxy_identification}, we select multiple galaxy candidates which are potentially associated with the absorbers. In order to minimize the uncertainty given by $z_{phot}$ in finding $the$ most likely absorber host galaxy, we measure a weighted average physical parameter (stellar mass, SFR, and impact parameter $D$) of all the galaxy candidates instead. The stellar mass, SFR, and luminosity contribution from a single galaxy candidate to the average parameter are weighted by their $z_{phot}$ likelihood distribution area under the $\pm$1000 km s$^{-1}$ velocity window at the absorber redshift. 

\begin{figure*}
  \resizebox{\hsize}{!}{\includegraphics{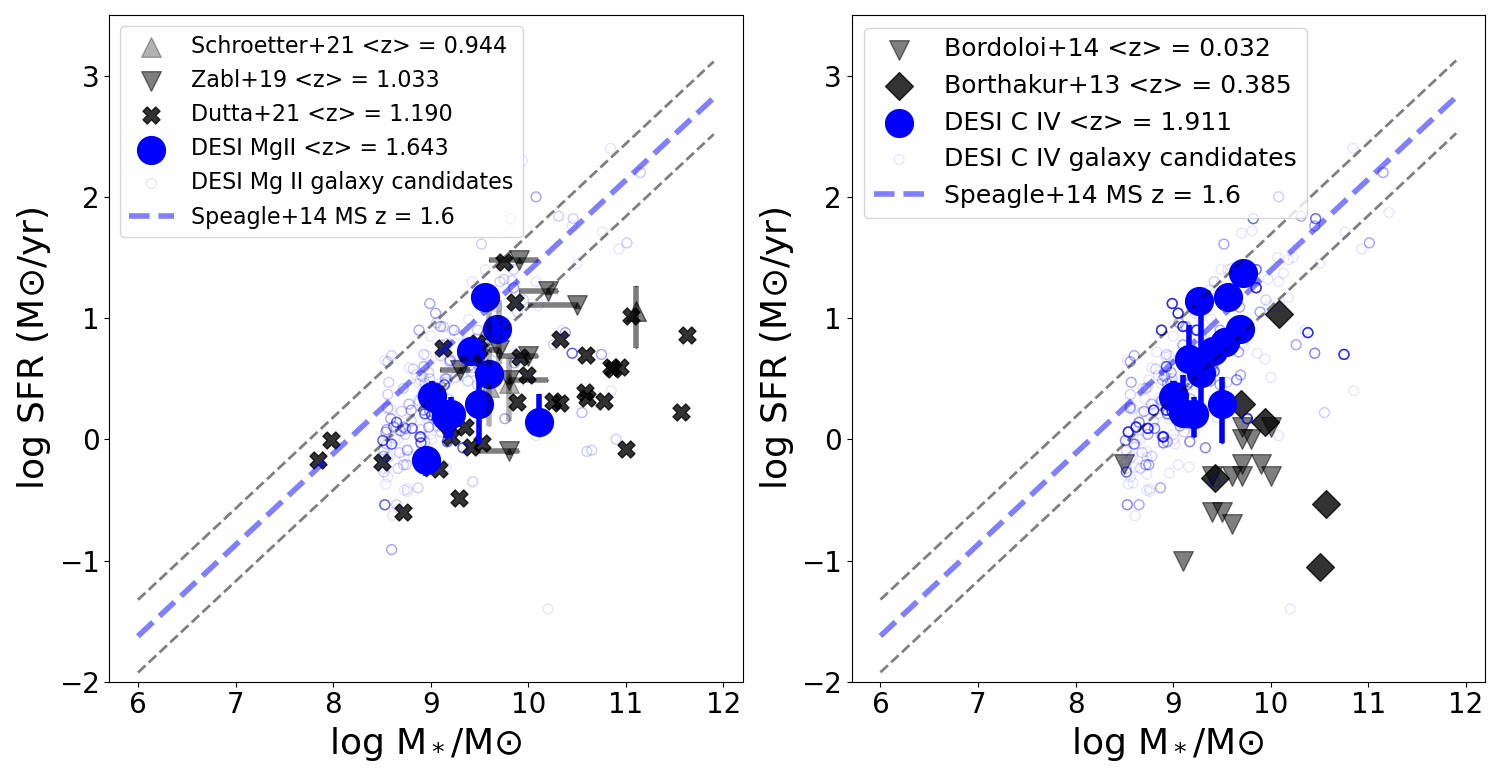}}
  \caption{\small{The SFR-$M_*$ relation of \MGII~(left) and \CIV~(right) host galaxies, respectively. The blue dashed line is the $M_*$-SFR relation of the main sequence galaxies at $z\sim$ 1.6 with 0.3 dex scatter in \citet{spea14}. The blue dots are the average values of \MGII~and \CIV~absorbing galaxies. Details of the method for obtaining the average values are described in Section \ref{sec:galaxy_identification}. The hollow circles are the pre-selected galaxy candidates. $Left$. The black crosses are the \MGII~galaxies from \citet{dutta20}. The upward and downward triangles are \MGII~galaxies in \citet{sch21} and \citet{zab19}. $Right$. The black diamonds are the \CIV~galaxies in \citet{bort13} and the grey triangles are the \CIV~galaxies in \citet{bor14}.
}}\label{fig:sfr_ste_mass_mg_civ}
\end{figure*}

In Figure \ref{fig:sfr_ste_mass_mg_civ}, we plot the $M_*$-SFR relation of \MGII~and \CIV~absorbing galaxies. The blue hollow circles represent all the selected galaxy candidates colored by their $z_{phot}$ distribution area probability within the velocity window of $\pm$1000 km s$^{-1}$ of the absorber. The filled blue circles are the weighted average parameters in each quasar field. The gray lines are the $M_*$-SFR relation of the main-sequence galaxies (MSG) at $z = 1.6$ from \citet{spea14} with 3 dexes of scatter. The pre-selected \MGII~and \CIV~galaxies have stellar masses ranging from 10$^{\sim8.5-11}$ M$_\odot$. Most of the weighted stellar masses are in the 10$^{9-10.5}$ M$_\odot$ range. 
 
\subsection{Correlation between gas and galaxy properties}

We then explore the dependence of absorbing gas strengths and ionization state on the host galaxy properties such as SFR, $M_*$, V-band luminosity $L_v$ and the impact parameter $D$. Empirical correlations between the \MGII~absorbing gas and host galaxy properties at $z<1$ have been reported in the literature, such as the extent of \MGII~absorbing gas $R_{gas}$ with the stellar mass relation \citep{chen10b}; the $R_{gas}$ with galaxy luminosity $L_B$ relation\citep{chen10}; the $W_r$ of \MGII~and the luminosity of [O~{\sc ii}] emission in the host galaxy relation by stacking the SDSS spectra \citep{men11} and the \MGII~$W_r$-$M_*$ relation \citep{bor11}. The $W_r$-$D$ anti-correlation is reported in extensive work (e.g., \citealt{nie13b,rud18a}), and the scatter in this relation is then explained by the segregation in the gas halo mass difference \citep{chur13a}.

\begin{figure*}
  \resizebox{\hsize}{!}{\includegraphics{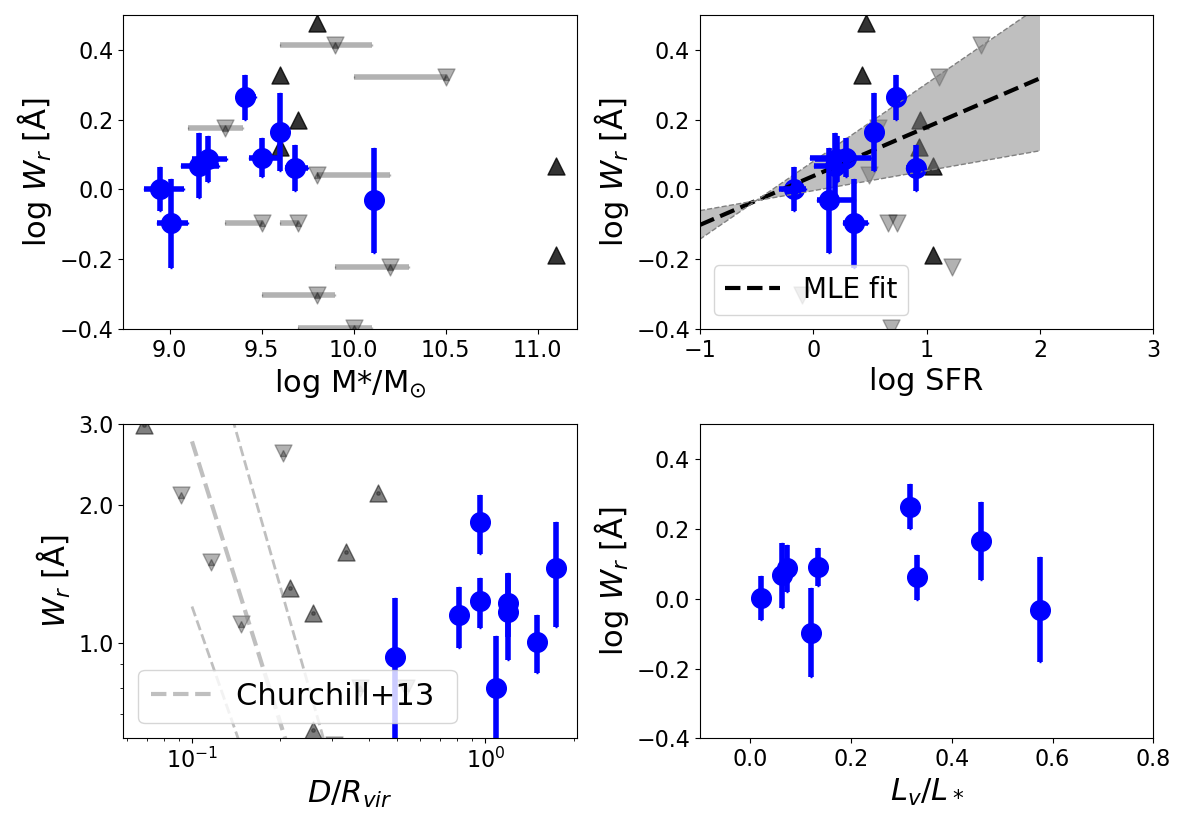}}
  \caption{\small{The \MGII~absorbers equivalent width dependence on the host galaxy properties. The labels for data in each survey are consistent with the labels in the left panel of Figure \ref{fig:sfr_ste_mass_mg_civ}. The upper panels are $W_r$-$M_*$, $W_r$-SFR relations and the lower panels are $W_r$-$D$/$R_{vir}$ and $W_r$-$L_v/L_*$ relations, respectively, indicative of a significant dependence of $W-r$ on the galaxy SFR. Different than what found in \citet{chur13a}, the $W-r$ of strong \MGII~at $z$ = 1--2 have a relatively constant dependence on the impact parameter $D$ normalized by the virial radius $R_{vir}$. Note that we do not include the system having $W-r$ - 0.28 \AA in this plot. 
}}\label{fig:mgii_galaxy}
\end{figure*}

\begin{figure*}
  \resizebox{\hsize}{!}{\includegraphics{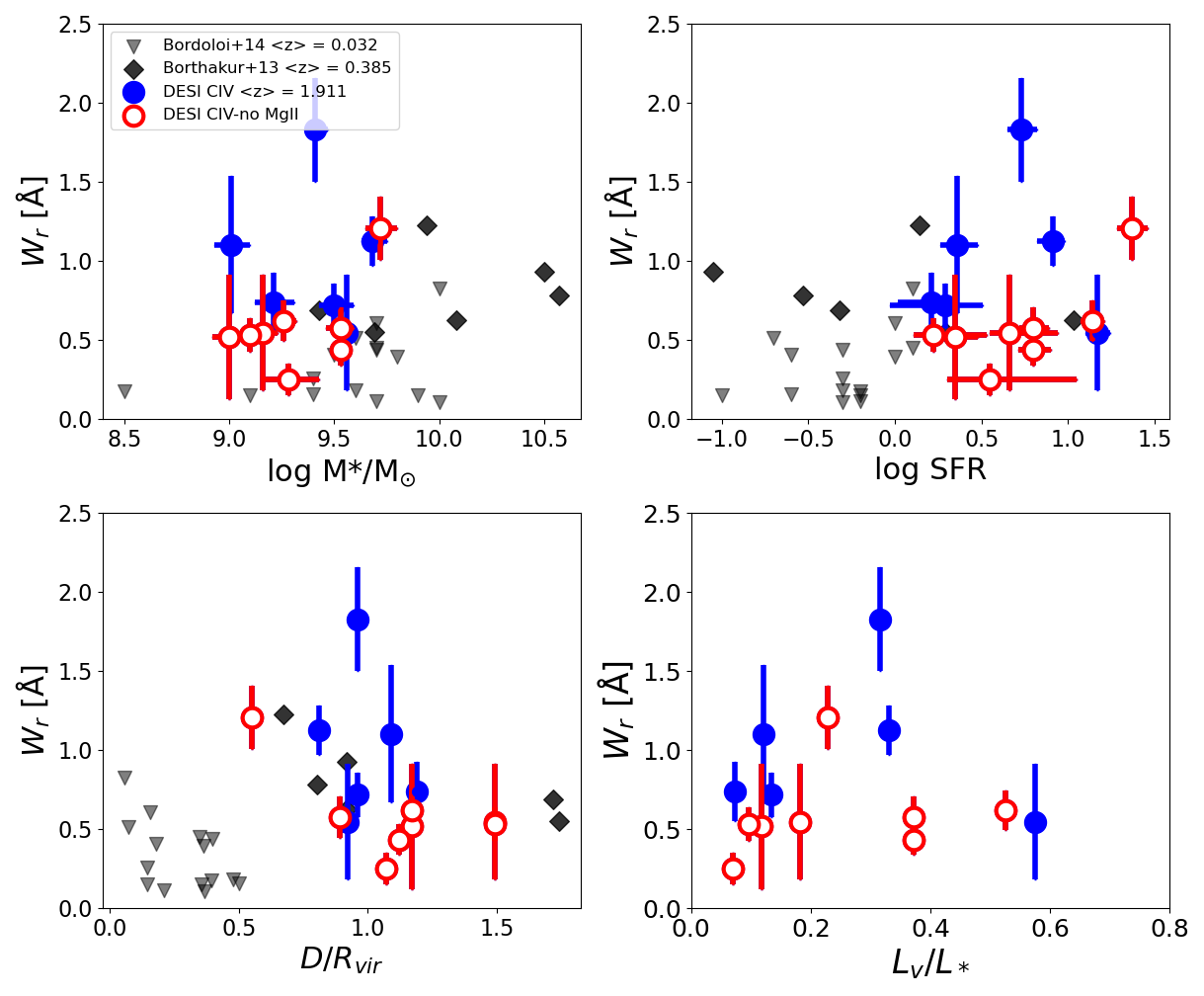}}
  \caption{\small{The \CIV~absorbers equivalent width dependence on the host galaxy properties. The labels for data in each survey are consistent with the labels in the right panel of Figure \ref{fig:sfr_ste_mass_mg_civ}. The red circles are the \CIV~systems without \MGII~absorption. The upper panels are $W_r$-$M_*$, $W_r$-SFR relations and the lower panels are $W_r$-$D$/$R_{vir}$ and $W_r$-$L_v/L_*$ relations, respectively. No significant correlation is found between the \CIV~strength and $M_*$, SFR, and $D/R_{vir}$. Particularly, nearly half of our \CIV~galaxy $D$ is beyond the virial radius.  
}}\label{fig:civ_galaxy}
\end{figure*}

In figure \ref{fig:mgii_galaxy} and \ref{fig:civ_galaxy}, we plot the weighted average $W_r$-$M_*$, $W_r$-$SFR$ , $W_r$-$L_v/L_\odot$ and $W_r$-$D/R_{vir}$ relations of \MGII~and \CIV~ absorber and its host galaxies. The $R_{vir}$ is the virial radius, defined by 200 times overdensity of the critical cosmic density $\rho_{crit}$: $R_{vir}$ = (M$_h$/(3/4$\pi$)200$\rho_{crit}$)$^{1/3}$. We calculate the virial radius by converting the stellar mass into the halo mass using the stellar-halo mass relation in \citet{gir20}. %The $L_*$ is the characteristic galaxy luminosity at $z\sim 2$ ($\sim 10^{10.38}$ M$_\odot$).
We present the values of the weighted average galaxy parameters in Table \ref{table:gqp-mgii} and \ref{table:gqp-civ} as well.

We note that for \MGII, a tighter correlation is seen between the absorber strength with the average galaxy SFR than its luminosity and stellar mass. We assume a power-law correlation between $W_r$ and SFR and $L/L_*$: log $W_r$(\MGII) = a$\times$log(SFR) + b1 and log $W_r$(\MGII) = a$\times$log($L/L_*$) + b2. The maximum likelihood estimation result is log $W_r$(\MGII) = 0.14$\times$log(SFR) + 0.038 (1 $\sigma$). Different than the works in the local universe, we do not see a significant correlation between the \MGII~$W_r$ against the stellar mass, $D/R_{vir}$ or luminosity within $D$ = 250 kpc. Furthermore, we detect several \MGII~systems at $D/R_{vir}>$ 1.

For \CIV, no significant correlations are seen between the overall \CIV~equivalent width and the galaxy candidates' physical parameters. Additionally, we note that the \CIV~(with \MGII) systems and \CIV-only systems host galaxies do not exhibit significant differences in $M_*$, SFR, and $L_v$. A marginal trend is seen that the \CIV~(no \MGII) systems tentatively reside in a larger impact parameter than \CIV~(with \MGII) systems. 

To further strengthen our results about the correlation between the absorber and its environment, we generate the posterior distribution $P(r|X$) of the Pearson correlation coefficient $r$ between log$W_r$-$M_*$, log$W_r$-log SFR, and log$W_r$-$L_v$, where $X$ is the galaxy dataset. We assume there are $n$ galaxies in one quasar field that are associated with the absorber. The $P(r|X) = \sum_{X_n} P(r|X_n)P(X_n|X)$. For one quasar field, we have $n$ galaxies. We perform a Markov chain Monte Carlo to sample the dataset. We present the result in Appendix Figure \ref{fig:pearson}. From the figure, we can tell that, for \MGII, the $W_r$ has a more significant correlation with SFR than stellar mass. For \CIV, no significant difference is seen between the $W_r$-SFR, $W_r$-$M_*$, and $W_r$-$L/L_{\odot}$ relation. This is consistent with the results when including the major galaxy candidates. This method is unbiased in selecting the absorbing host galaxies with a velocity window and considering the $z_{phot}$ uncertainties in a statistical way.

%We also detect other metal lines such as Si~{\sc iv}, Al~{\sc ii}, Al~{\sc iii}, and Fe~{\sc ii} in the 14 sightlines in the COSMOS field. The rise of intermediate ions Si~{\sc iv} and C~{\sc iv} are connected to the phoionization and collision state of the warm gas and/or the transition phase \citep{werk19}. The ratio of N(Si~{\sc iv})/N(C~{\sc iv}) is a possible indicator of the fraction of C~{\sc iv} in the total carbon species content, i.e. the ionization state of the gas. Therefore, in Figure \ref{fig:ionization}, we plot the ionization parameter indicated by the equivalent width ratio of high-ion C~{\sc iv} and low-ion Si~{\sc iv} as a function of $D/R_{vir}$. It is clear that the multiphase gas in our sample extends out to twice the virial radius. The ionization parameter increases significantly out of the virial radius. 

%\begin{figure}
% \resizebox{\hsize}{!}{\includegraphics{ionization_civ_siiv.png}}
%   \caption{\small{The equivalent widths ratio of high-ion C~{\sc iv} and low-ion Si~{\sc iv} as a %function of $D/R_{vir}$. The ionization parameter indicated by the $W_r$(\CIV) / $W_r$(Si~{\sc iv}) of the CGM increases with the $D/R_{vir}$.  
%}}\label{fig:ionization}
%\end{figure}

\subsection{Covering fraction of \MGII~and \CIV~absorbing gas}

%%%%%%%%%%%%%%%%%%%%%%%%%%% Definition of the fc%%%%%%%%%%%%%%%%%%%%%%%%%%%%%%%%%%%%%%%%%%%%%%%%%%%%%%%%%

Here we discuss the absorbing galaxy fraction among different galaxy populations above a certain detection limit. The covering fraction $f_c$ within an impact parameter $D$ is defined as the probability of a galaxy having a metal absorption line ($W_r > W_{lim}$) that can be detected. The probability can be calculated by the ratio of absorbing galaxy number versus the total galaxy number within a certain $D$ \citep{bor14}: \[f_c(W_r > W_{lim}) = \frac{N_{abs}}{N_{tot}}\], for which $N_{abs}$ is the number of galaxies having $W_r> W_{lim}$ galaxy number and $N_{tot}$ is the total galaxy number in the same field. We use the same method to select the potential absorbing galaxies that are discussed in Section \ref{sec:galaxy_identification}.

\begin{figure*}
 \resizebox{\hsize}{!}{\includegraphics{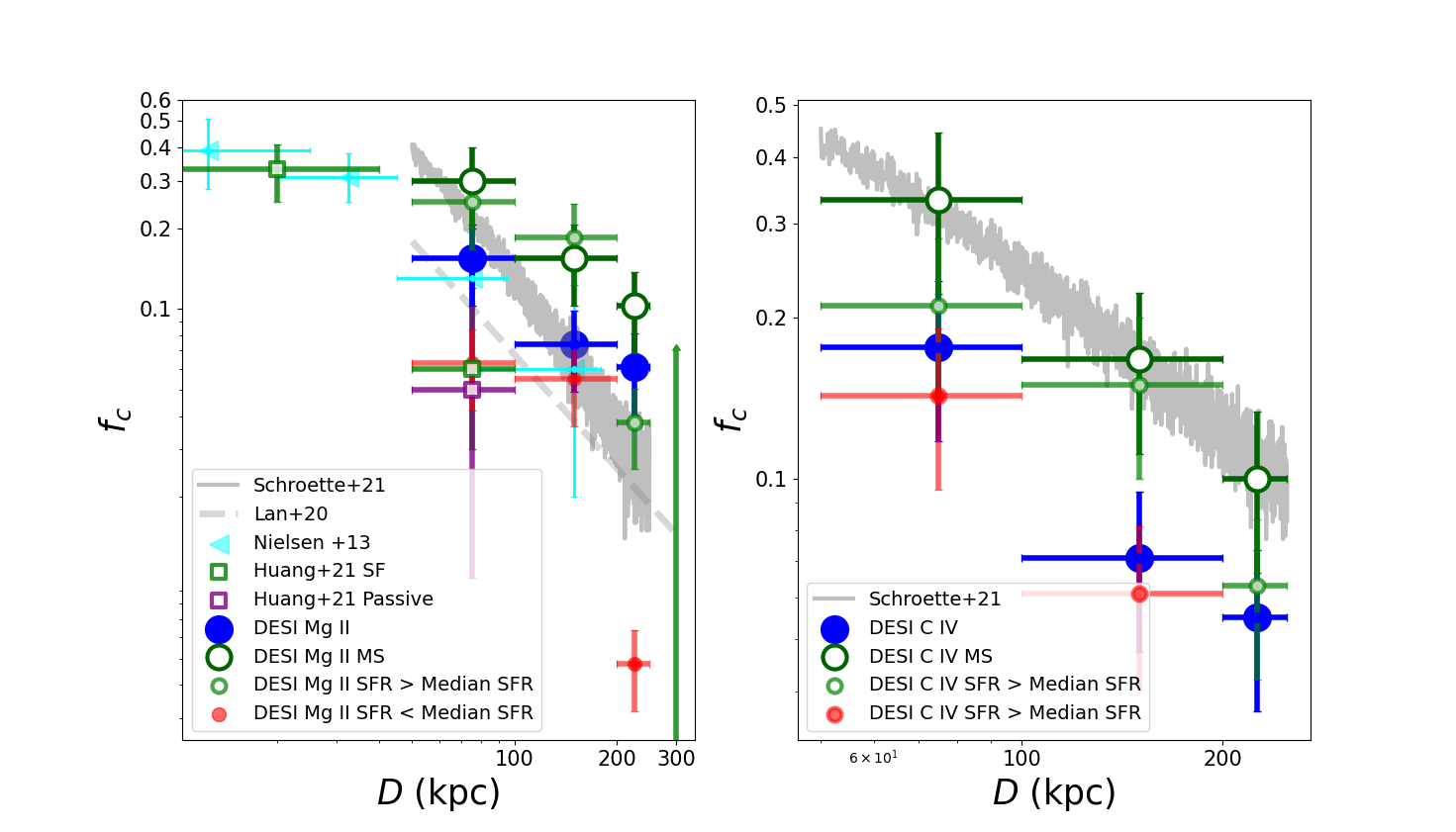}}
   \caption{\small{The covering fraction of \MGII~($z$ = 0.9 -- 2.1, $W_r\geq$ 1.0 \AA) and \CIV~absorbing gas ($z$ = 1.5 -- 2.1, $W_r\geq$ 0.5 \AA) in all the galaxies (blue) and main-sequence galaxies (dark green) as a function of impact parameter $D$ within 300 kpc. We also plot the covering fraction of absorbing galaxies in the subsamples with SFRs greater than the median SFR (smaller green circles) and those with SFRs less than the median SFR (red circles). The results of this work are calculated in three bins of $D$: 50-100 kpc, 100-200 kpc, 200-250 kpc. The green and purple squares in the left panel are the $f_c$ of \MGII ($W_r >$ 1.0 \AA) in star-forming and passive galaxies in \citet{huang21} and \citet{chen10}. The grey curves are the $f_c$ distribution as a function (95 \% confidence region) of $D$ in \citet{sch21} with a mean \MGII~and \CIV~equivalent width of 0.99 and 0.70 \AA respectively. The cyan and triangles are the $f_c$ of strong \MGII~($>$ 1\AA) absorbing gas in the MAGIICAT survey at $z<1$ \citet{nie13a}. The grey dashed line is the $f_c$-D relation of strong \MGII~($>$ 1\AA) absorbing gas in \citet{lan20}. 
}}\label{fig:fc_mg_civ}
\end{figure*}

%%%%%%%%%%%%%%%%%%%%%%%%%% Compare fc-D relation with MAGAFLOW and z < 1 surveys%%%%%%%%%%%%%%%%%%%%%%%%%%%%%%%%%%%%%%%
In our case, we count the number of absorbing galaxies weighted by the $z_{pdf}$ distribution within the velocity window $p_z$ of each galaxy. The median $p_z$ in each quasar field is dominated by the galaxies that are most correlated with the absorber. We neglect the galaxies whose $p_z$ is 1 dex smaller than the major galaxi(es).
We divide the differential \MGII~and \CIV~absorbing galaxies covering fraction at 0.9 $<z<$ 2.2 into three impact parameter bins (50-100, 100-200, and 200-250 kpc). The covering fraction is plotted in Figure \ref{fig:fc_mg_civ}. We find that within 250 kpc, an anti-correlation between the $f_c$ and the $D$ is seen for both \MGII~and \CIV. In Section \ref{sec:gas_galaxy}, we present that the strength of our \MGII~absorption has the most significant dependence on the SFR of host galaxies, therefore, we check the $f_c$ of strong \MGII~in the main sequence galaxies, $f_c(MSG)$ =  $N_{MgII}/N_{MSG}$. We select the main sequence galaxies based on the relation in \citet{spea14} within 0.3 dex. The $z_{pdf}$ of each galaxy is used for the target selection. We find that the strong \MGII~covering fraction in main-sequence galaxies (dark green circles) is two times higher (0.30 within 100 kpc) than that in the whole galaxy population (blue circles) in Figure \ref{fig:fc_mg_civ}. Given the arbitrariness of the definition of main sequence galaxies within a 0.3 dex scatter and the uncertainties associated with estimating the main sequence, we further include two subsamples of galaxies with SFRs greater and smaller than the median SFR (green and red circles in Figure \ref{fig:fc_mg_civ}). The $f_c$ value of \MGII~galaxies in the subsample with SFR $>$ median SFR is similar to $f_c(MSG)$ and exhibits a significant evolution compared to those in the local star-forming galaxies in \citet{huang21}.

\begin{table*}
\centering
  \caption{\MGII-absorbing galaxy candidates information \label{table:gqp-mgii} }
  
\begin{tabular}{llllllllll}
\hline
QSO & RA & DEC & $D$ & $D/R_{vir}$ & $z_{abs}$&$W_r$ & log(SFR) & log M*/M$_{\odot}$& $L_v/L_\odot$\\
&&&(kpc)&&&(\AA)&(M$_*$/yr)&& \\
%  &($W_r>$ 0.3 \AA) &($W_r>$ 1.0 \AA) & & ($W_r>$ 0.2 \AA) &($W_r>$ 1.0 \AA) \\
\hline
\small
%39627835572226082  
J100219.49+015536.84 & 150.5812 & 1.9269 & 218.53 & 1.79 & 0.9520 &  1.462$\pm$0.316   & 0.54$^{0.07}_{-0.07}$   &   9.6$^{0.04}_{-0.03}$     &   10.04    \\
%39627835572226082  
J100219.49+015536.84 & 150.5812 & 1.9269 & 102.62  & 0.31 & 1.2090 &  0.932$\pm$0.260   & 0.0$^{0.068}_{-0.068}$   &  10.11$^{0.04}_{-0.03}$   &   10.14     \\
%39627853674842755  
J095834.03+024426.88 & 149.6418 & 2.7408 & 181.26 & 2.21 & 1.2756 &  1.005$\pm$0.126   & --0.17$^{0.072}_{-0.297}$ &   8.95$^{0.09}_{-0.13}$    &    8.7     \\
%39627823471662212  
J095749.98+013354.10 & 149.4582 & 1.5650 &   144.51 &  0.73 & 1.5660 &  1.169$\pm$0.202   & 0.19$^{0.18}_{-0.13}$   &   9.16$^{0.10}_{-0.11}$    &   9.18      \\
%39627823471662212 
J095749.98+013354.10  & 149.4582 & 1.5650 &  153.39 &  0.73 & 1.6428 &  1.223$\pm$0.162  & 0.21$^{0.19}_{-0.14}$  &   9.21$^{0.09}_{-0.10}$    &   9.24      \\
%39627829524040255  
J100031.61+014757.48  & 150.1317 & 1.7993 &  150.82 & 1.89 & 1.6625 &  1.836$\pm$0.257  & 0.73$^{0.08}_{-0.09}$   &   9.41$^{0.05}_{0.06}$     &    9.88      \\
%39627835559645239  
J095949.39+020140.80 & 149.9558 & 2.0280 &   137.37 & 0.74 & 1.7372 &  1.152$\pm$0.175   & 0.91$^{0.09}_{-0.07}$   &  9.68$^{0.05}_{-0.07}$    &     9.9     \\
%39627823471662212
J095749.98+013354.10 & 149.4582 & 1.5650 &  146.41 & 0.43 & 1.8770 &  1.234$\pm$0.114   & 0.29$^{1.262}_{0.829}$   &  9.50$^{0.07}_{-0.09}$    &     9.51      \\
%39627835563836564  
J100014.14+020054.36 & 150.0589 & 2.0151  & 149.83 & 1.85  &  1.9813 &  0.300$\pm$0.117 & 1.17$^{0.32}_{-0.22}$   &   9.56$^{0.04}_{-0.04}$    &    10.14      \\
%39627841607828079  
J100105.30+021348.00 & 150.2703  &  2.2312  &   66.70  & 0.48 & 2.1680   &  0.800$\pm$0.235  & 
0.36$^{0.1 }_{0.12 }$   & 9.01$^{0.08}_{0.09}$ &  9.46  \\   
\hline
\end{tabular}
\end{table*}

\begin{table*}
\centering
  \caption{\CIV-absorbing galaxy candidates information \label{table:gqp-civ} }
\begin{tabular}{llllllllll}
\hline
QSO & RA &  DEC & $D$ & $D/R_{vir}$ & $z_{abs}$&$W_r$& log(SFR) & log M*/M$_{\odot}$& $Lv/L_\odot$ \\
&&&(kpc)&&&(\AA)&(M$_*$/yr)&& \\
%  &($W_r>$ 0.3 \AA) &($W_r>$ 1.0 \AA) & & ($W_r>$ 0.2 \AA) &($W_r>$ 1.0 \AA) \\
\hline
\small
%39627823471662212
J095749.98+013354.1 & 149.4582 & 1.5650   &   150.82  & 0.73 & 1.6428   &  0.738$\pm$0.200    & 0.21$^{0.19}_{-0.14}$ &  9.24$^{9.37 }_{9.15 }$ & 9.24  \\ 

%39627829524040255 
J100031.61+014757.48  & 150.1317 & 1.7993 &  129.28   & 1.97 & 1.6625   &  1.828$\pm$0.270   &  0.73$^{0.08}_{-0.09}$  & 9.88$^{9.66 }_{9.55 }$ & 9.88 \\

%39627835559645239
J095949.39+020140.80 & 149.9558 & 2.0280  &   137.37  & 1.87 & 1.7372   &  1.126$\pm$0.080    & 0.91$^{0.09}_{-0.07}$  & 9.9$^{10.16}_{10.03}$&  9.9 \\ 

%39627829536620822
J100302.90+015208.40 & 150.7621 &  1.8690 &   178.71 & 0.75   & 1.7970 &  0.531$\pm$0.104     &  0.22$^{0.11}_{-0.31}$ & 9.36$^{8.83}_{8.6}$  &  9.36 \\ 

%39627835563836564  
J100014.14+020054.36 & 150.0589 & 2.0151   &  137.45  & 1.83 & 1.8400   &  0.574$\pm$0.147  & 0.8$^{0.25}_{-0.5}$  & 9.95$^{8.23 }_{7.94 }$ &  9.95 \\

%39627853674842755
J095834.03+024426.88 &  149.6418 & 2.7408  &  142.13   & 1.30 & 1.8555   &  0.252$\pm$0.100   & 0.55$^{0.25}_{-0.5}$  &  9.22$^{9.12 }_{8.81 }$ &  9.22 \\ 

%39627823471662212
J095749.98+013354.10 & 149.4582 & 1.5650  &    146.41  & 2.33 & 1.8770   &  0.716$\pm$0.142   &   0.29$^{0.32}_{-0.22}$  &
9.51$^{10.54}_{10.38}$  &  9.51\\
%39627835563836564
J100014.14+020054.36 &  150.0589 & 2.0151  &   152.60  & 0.46 & 1.9450   &  0.620$\pm$0.099   &  1.14$^{0.07}_{0.07}$ & 10.1$^{9.12 }_{8.97 }$  &  10.1 \\ 

%39627841591054723 
J095752.30+022021.12 & 149.4679 &  2.3392  &   177.02  & 0.46 & 1.9810   &  0.546$\pm$0.200   &  0.66$^{0.11}_{-0.28}$  & 9.64$^{8.87 }_{8.62 }$ &  9.64 \\ 

%39627835563836564
J100014.14+020054.36 & 150.0589 & 2.0151   &   149.84  & 1.83 & 1.9813   &  0.453$\pm$0.114    & 1.17$^{0.07}_{-0.07}$  &
10.14$^{9.12 }_{8.97 }$ &  10.14 \\
%39627835563836564 
J100014.14+020054.36 & 150.0589 & 2.0151   &   135.46  & 2.19 & 2.1305   &  0.434$\pm$0.071    & 0.80$^{0.09}_{0.1}$  & 9.95$^{11.08}_{10.98}$ &  9.95 \\

%39627841607828079 
J100105.30+021348.00  & 150.2721  & 2.2300  &  133.99  & 0.49 & 2.1530   &  0.518$\pm$0.341  & 0.35$^{0.1}_{-0.13}$ & 9.45$^{9.432}_{9.314}$ &  9.45  \\

J100105.30+021348.00  & 150.2721  &  2.2300 &  137.10   & 0.71 & 2.1680   &  1.102$\pm$0.221 &  0.36$^{0.1}_{-0.12}$ & 9.46$^{9.432}_{9.314}$ &  9.46  \\

%2937
J095806.96+022248.36  & 149.5209  &  2.3814  &  117.77   & 0.42 & 3.0880   &  1.207$\pm$0.211  & 1.37$^{0.09 }_{-0.09}$  & 9.74$^{8.74 }_{8.67 }$ &  9.74 \\ 

%39627823471662212,1.6428
%39627829524040255,1.6625
%39627835559645239,1.7375
%39627829536620822,1.797 
%39627835563836564,1.84  
%39627853674842755,1.8555
%39627823471662212,1.877 
%39627835563836564,1.945 
%39627841591054723,1.981 
%39627835563836564,1.9813
%39627835563836564,2.1305
%39627841607828079,2.153 
%39627841607828079,2.168 
%39627847635042937,3.088
  
 \hline
\end{tabular}
\end{table*}

\subsection{Comparison with other surveys}

We compare our CGM-galaxy correlation results with other surveys in this section. We are cautious about comparing surveys having different depths and sample selection methods. Therefore we only discuss the trends rather than the quantitative differences here. 

For \MGII, we compare our results with the strong \MGII-galaxy correlation in the MAGIICAT \citep{nie13a}, MAGAFLOW survey \citep{sch19,zab19} and the DECaLS imaging survey \citep{lan20}. The galaxies selected in MAGIICAT have a similar depth as ours (B-band magnitude limit of -16.1, i.e., around 24.5 mag at $z = $ 0.34). The DECaLS imagings used in \citet{lan20} have $g$ and $z$ band limits of 24.2 and 23 mag, respectively. We perform the Kolmogorov–Smirnov test between the $M_*$-SFR relation of the sample and $M_*$-SFR of galaxies in the MAGAFLOW inflowing-mode (upward triangles) and outflowing-mode (downward triangles). The $p$-values are 0.014, and 0.0153, respectively, suggesting that our \MGII~gas does not exhibit obvious inflowing or outflowing features. In other words, the strong \MGII~gas at $z =$ 1--2 tentatively exhibits complex kinematics, which may be a combination of effects by gas accretion, galactic outflows and gas recycling, etc. 

We compare the \MGII~having $W_r > $ 1 \AA~ gas $f_c$ with that at $z=$ 1--1.5 \citep{lan20} (the grey dashed line and grey line, respectively) and $z<1$ \citep{nie13b} (grey triangles). We note that our $f_c$ of our strong \MGII~absorbing gas shows a marginal evolution than that at $z<1$ \citep{nie13b}. The \MGII-absorbing gas covering fraction in either main-sequence galaxies or star-forming galaxies has a significant evoulution from $z = 0$ to $z =$ 1--2.5. \citet{lan20} also find that the covering fraction of strong \MGII~systems in star-forming galaxies is higher than that in passive galaxies, exhibiting a significant evolution from $z$ = 0.4 to $z$ = 1.3. %We note that our $f_c$ value is higher than that in \citet{lan20} can be due to the COSMOS survey depth being deeper than the DECaLS survey. 
\citet{sch21} provide a novel Hamiltonian Monte Carlo model optimizing for estimating the $f_c$ with limited samples. Their $f_c$ is for \MGII~systems having $W_r>$ 0.6 \AA, it is clear that our strong \MGII~gas $f_c$ is higher than their $f_c$ beyond 100 kpc.

We compare our \CIV~$M_*$-SFR relation with \CIV~host galaxy from the COS-Dwarfs \citep{bor14a} and COS-Halo surveys \citep{bort13} at $z<1$. %and the KBSS survey of 135 $z\sim$ 2 star-forming galaxies \citet{rud19} at 2.1$< z <$ 2.7. 
\citet{bor14a} study a sample of 43 sub-$L_*$/dwarf galaxies at $z<$ 0.1. A tentative correlation between the \CIV~$W_r$ and host galaxy SFR has been reported in \citet{bor14a} at $z<1$. \citet{bort13} study 20 galaxies at $z<$ 0.2 and find a high detection rate of \CIV~systems in starburst galaxies. The detection rate of \CIV~is 4/5 in starburst galaxies(blue triangles in Fig. \ref{fig:civ_galaxy}) and 2/12 in the control sample comprising of normal/passive galaxies (grey triangles). 

We note that our \CIV~$f_c$, unlike the strong \MGII~gas, is lower than that in the MAGAFLOW survey \citet{sch21} at $z = $1--1.5 within 250 kpc. This may be because that half of our \CIV~systems are \MGII~bearing halos, and the galaxies associated with weaker \CIV-only halos are underestimated. In \citet{sch21}, the authors suggest that at $z$ = 1--1.5, the covering fraction of \CIV-only gas is larger than that of \MGII~and \CIV gas within 250 kpc and it likely resides a broader radius in the IGM (out to 250 kpc). 

\begin{table*}
\centering
  \caption{The covering fraction of \MGII~and \CIV in all the galaxies and main-sequence galaxies at different impact parameter bins (50-100, 100-200, 200-250 kpc) respectively. The galaxies are searched within 30 arcseconds offset by the quasar(absorbing system) using the COSMOS2020 catalog. \label{table:fc_mgii_civ} }
\begin{tabular}{ccccccc}
\hline
$D$  & $f_c$ \MGII  & $f_c$ \MGII~(MSG) & $f_c$ \CIV & $f_c$ \CIV~(MSG)  \\
\hline
 50 -- 100 kpc    & 0.152 $\pm$ 0.076 & 0.290 $\pm$ 0.127  & 0.176 $\pm$ 0.024  & 0.333 $\pm$ 0.083    \\ 
 100 -- 200 kpc    & 0.074 $\pm$ 0.119 & 0.154 $\pm$ 0.293  & 0.071 $\pm$ 0.022  & 0.167 $\pm$ 0.051     \\    
 200 -- 250 kpc    & 0.061 $\pm$ 0.187 & 0.103 $\pm$ 0.141  & 0.055 $\pm$ 0.017  & 0.100 $\pm$ 0.038     \\ 

\hline
\end{tabular}
\end{table*}

%%%%%%%%%%%%%%%%%%%%%%Correlation with D/Rvir, i.e. take the halo mass into consideration, %%%%%%%%%%%%%%%%%%%%%%%%%%%%%%%%%%%%%%%%%

\section{Discussion}\label{sec:discussion}

\subsection{Strong \MGII~absorbing gas contribution to global SFR at $z$ = 1--2}
We follow the method described in \citet{nes11} to estimate the strong \MGII~gas contribution to the global SFR density: \[F  = \rho^{MgII}_{SFR}/\rho_{SFR} = <SFR> \times \frac{dN}{dz}\frac{dl}{dz}^{-1}\sigma^{-1}/\rho_{SFR}\], where $<SFR>$ is the mean SFR of \MGII~host galaxies, the dN/dz is the line density of \MGII~absorbers, $dl/dz$ is the differential proper distance, $\sigma$ is the \MGII~gas cross-section and the $\rho_{SFR}$ is the global star formation density. We take our detected strong \MGII~galaxy $<SFR>$ of 7.44 M$_{\odot}$ yr$^{-1}$, the dN/dz at 1.5 -- 2.0 = 0.753 $\pm$ 0.141, the cross-section of \MGII~gas as (250 kpc)$^2$. The $\rho_{SFR}$ at $z =$ 2.0 calculated from Equation 15 in \citet{madau14} is 0.132 M$_{\odot}$ yr$^{-1}$ Mpc$^{-3}$. We then estimate the contribution of strong \MGII~absorbing gas to the global SFR density at $z\sim$ 2.0 is 0.068 by assuming a gas cross-section of (250 kpc)$^2$. This fraction is consistent with our measured covering fraction of strong \MGII~in star forming galaxy subsample within 250 kpc, further suggesting the co-evolution of cool gas probed by the strong \MGII~systems and the cosmic star formation activity.

%{\bf We estimate the cross-section of \MGII~absorbing star-forming galaxies from its covering fraction. The median SFR for the star-forming galaxies subsample is 6.72 M$_{\odot}$ yr$^{-1}$. In \citep{lan20}, they took an assumption of \MGII~gas cross-section in galaxies at $z\sim$ 0.7 is 6000 kpc$^2$. If we adopt our measured dN/dz of strong \MGII at 1.0 $<z<$ 2.5, we can estimate the corss-section of strong \MGII~gas   }
 
\subsection{\MGII~and \CIV~gas origins}
Here we discuss how the \MGII~and \CIV~absorbing gas fuel the galaxy star formation and are affected by the galactic feedback towards the comic noon. According to the results above, we find that two gas phases co-exist in the CGM in this work: one is with detectable \MGII~and \CIV~absorption in the same system, i.e., same physical position and shares similar kinematics profiles, which probes a relatively higher $N$(H~{\sc i}); one is with the \CIV-only systems, probing a lower $N$(H~{\sc i}) gas and a tentative larger impact parameter. Such multiphase structure of the CGM is clearly seen in the hydrodynamic simulations \citep{ford13,sur17}. In \citet{ford13}, the simulation reveals that within the 300 kpc of galaxies at $z=$ 0.25, the multiphase CGM generally probes a 10$^{4-4.5}$K photo-ionized gas. Low ions such as \MGII~and Si~{\sc iv} probe a denser gas and are closer to galaxies, while \CIV~can be associated with the cool gas in the inner region or the collisionally dominated highly ionized gas in a broader region. In this work, we mainly consider the CGM within 250 kpc of the galaxy, the gas mainly exhibits “halo fountains” region properties suggested by the simulation of \cite{oppe08}. This region is first fueled by the metal-poor inflowing gas but later enriched by the metal-rich and momentum-driven outflows, where the gas has complex kinematics. Recent TNG simulation has the resolution to resolve small-scale structures. The \MGII~halos are found to be highly structured, clumpy, asymmetric \citep{nel20} and with a variety of kinematics \citep{def20}. This likely explains why our \MGII~absorbing gas does not exhibit obvious inflowing or outflowing features. 

We find a tight correlation between the strong \MGII~equivalent width and the host galaxy SFR at $z =$1--2, suggesting the co-evolution of strong \MGII~absorbing gas and the main-sequence galaxies towards the cosmic noon. The conclusion is strengthened by the fact that the covering fraction of \MGII~absorbing gas in the main-sequence galaxies is two times higher than the covering fraction in all the galaxies. Particularly, we detect several \MGII~and \CIV~systems having D/R$_{vir}>1$, indicating that the cool gas still exists out of the virial radius. These unbounded gas may be driven by the star formation activity at the cosmic noon. In \citet{rud19}, they found that 70\% of their galaxies have some unbounded metal-enriched gas, suggesting galactic winds may commonly eject gas from halos at $z\sim 2$.

The \CIV~only systems, as suggested in the COS-Halos survey in \cite{bort13} and \citet{bor14}, likely trace the dwarf galaxies and/or starburst galaxies. The sub-L$_*$ galaxies undergo extended bursty star formation rather than the continuous 'normal' star formation in the super $L_*$ galaxies. Though we do not find the host galaxies exhibit significant differences between the \CIV-only systems and the \CIV(with \MGII) systems. This may be due to that our galaxy properties are weighted average values, representing a feature of multiple star-forming galaxies. 

\subsection{Environmental effects}

Environmental effects can also give rise to the optically thick cool gas. Environmental effects of the \MGII~absorbing gas are detected in the MAGG survey at $z\sim$ 1--1.5 \citep{dutta20,dutta21} and the MAGIICAT survey $z<$ 1 \citep{nie18,huang21}. Particularly, \citet{nie18} point out that the covering fraction and median $W_r$ of \MGII~absorbing gas residing in a group environment is larger than that of an isolated galaxy. The group environment kinematics are with more power for high-velocity dispersion similar to outflow kinematics. \citet{loft23} study the correlation between the log $N$(H~{\sc i})/cm$^{2}$ $>$ 19 absorbing gas and Ly$\alpha$ emitters at $z = $ 3--4 and find that the optically thick gas covering fraction in galaxy group is three times higher than that in isolated systems.

In \citet{chen10} and \citet{huang21}, the authors find that the \MGII~gas that does not show obvious anti-correlation with $D$ resides in a group environment rather than associated with a single galaxy. Thus, the environmental effect is also a plausible explanation of our \MGII~and \CIV~systems detected out of the virial radius, and no obvious anti-correlation is seen in our \MGII~$W_f$ with $D/R_{vir}$. Additionally, the \CIV~system with the largest equivalent width has three major components in the absorption profile of \CIV, \MGII~and Si~{\sc iv} absorption. This system also has a large $D/R_{vir}$ value (1.97), which may indicate it is affected by the outflows or reside in a group or disturbing environment.

\subsection{Absorbing gas and non-absorbing gas counterparts}

\begin{figure}
  \resizebox{\hsize}{!}{\includegraphics{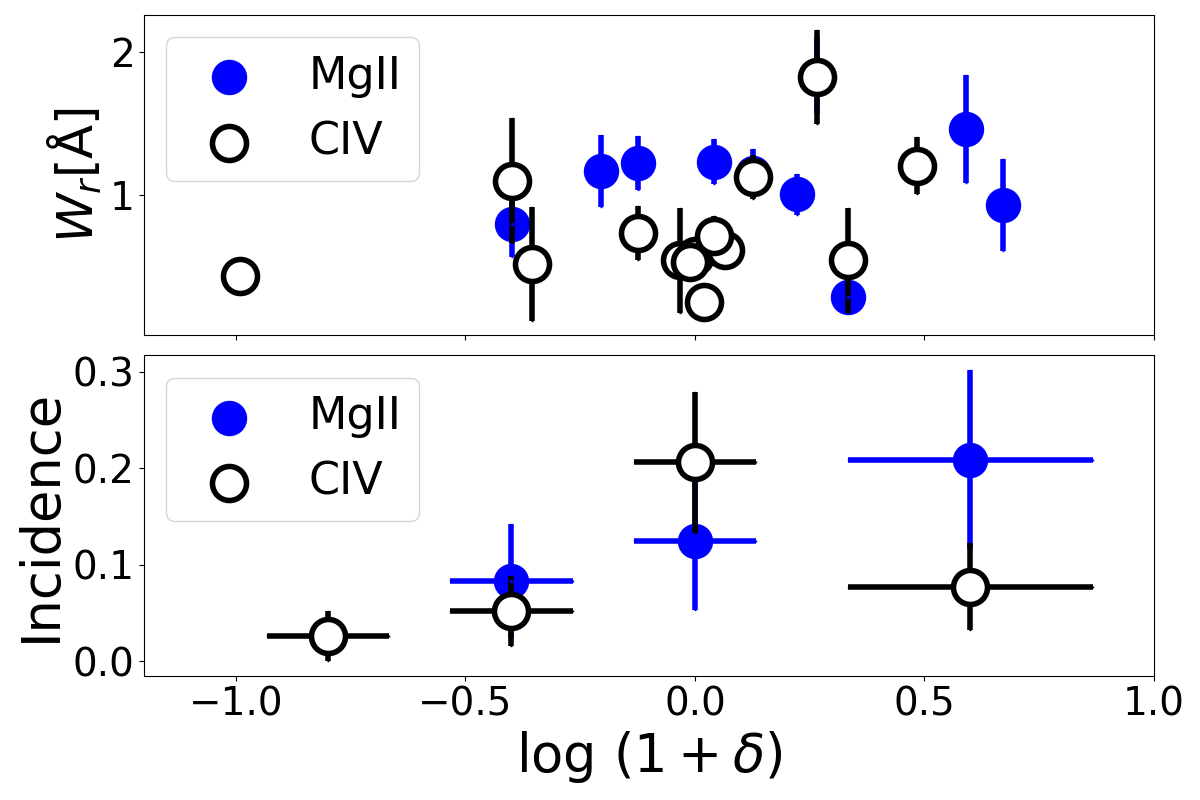}}
\caption{\small{The cross-correlation between \MGII~and \CIV~incidence and equivalent width with the galaxy number overdensity $\delta$. The $p$ value for \MGII~and \CIV~incidence with galaxy number overdenstiy are 0.047 and 0.590, 
}}\label{fig:cross-correlation}
\end{figure}

In order to further test if there is an intrinsic correlation between the absorbing gas with the galaxy overdensity, we compare the galaxy density at the redshift of absorbers and non-absorbers. In Appendix Figure \ref{fig:g_density}, we plot the normalized galaxy $z_{phot}$ density P($z$) at 0.0 $<z<$ 6.0. The \MGII~and \CIV~absorber redshifts are labeled by red and blue lines, respectively. The P($z$) is a normalized photo-$z$ probability distribution by taking around all the galaxies in the COSMOS2020 catalog within 30 $\arcsec$ offset the quasar sightline. We find that the \MGII~and \MGII(with \CIV) absorbers likely occur at the redshift where the galaxy density is higher than the redshifts where there are no absorbers between $z = 1 - 2.5$. The \CIV-only systems, i.e., the relatively lower $N$~(H~{\sc i}) systems, exhibit a mild correlation with the galaxy density. We carry out a cross-correlation analysis between the incidences of \MGII~and \CIV~absorbers, and their rest-frame equivalent width $W_r$, with galaxy overdensity. The methodology is similar as which presented in \citet{dutta21}. The galaxy overdensity, denoted as $\delta$, is defined by the relation $\delta = (\rho_{abs} - \rho_0)/\rho_0$. Here, $\rho_{abs}$ and $\rho_0$ are the absorbing-related and field galaxy number density, respectively. We calculate the $\rho_{abs}$ within a velocity window of $\pm$1000 km/s centered on the absorber redshif. The projected area is defined using an annulus with a radius of 250 kpc. The field galaxy number density in a cube volume of 300$\times$300 kpc and d$z$ = 0.2. %The cross-correlation function between two variables are described by a power-law: $\xi(r) =(r/r_0)^{-\gamma}$, where $r_0$ is the comoving distance at which the local number density of galaxies is twice that in an average place in the Universe, and $\gamma$ is the slope parameter. 
%In stead of three-dimensional cross-correlation function, we instead use the cross-correlation function in a projected radius $p_r$. 
We plot the relation between \MGII~and \CIV~$W_r$ and incidence with the galaxy number overdensity in Figure \ref{fig:cross-correlation}. From the figure we can tell that the galaxy density is higher when the \MGII~absorber incidence goes higher. We perform a Pearson correlation analysis between \MGII~and \CIV~incidence between the galaxy overdesnity. The $p$ value for \MGII~and \CIV~incidence with galaxy number overdenstiy are 0.047 and 0.590, respectively. The $p$ value for \MGII~and \CIV~$W_r$ with galaxy number overdenstiy are 0.838 and 0.175, respectively. This tentatively suggest that the environmental effects indeed play an active role in the origin of strong \MGII~absorbing gas at the cosmic noon. We are somewhat cautious about drawing firm conclusions given the limited sample size. %Furthermore, the warmer gas may occur first in a lower galaxy density region, and the \MGII~traced cool gas is generated in the following galaxy assembly process until the central/major galaxy forms. 
The galaxy density in the large-scale structure may play a more significant role in the birth and evolution of the multiphase CGM than the specific 'host' galaxy properties. 

%{\bf I am a bit not sure how you determine both galaxies are MgII-associated..}, %{\bf also see References, because I think the conclusion may be a bit strong}.

\section{Summary}\label{sec:summary}
1. We detect 51 \MGII~and 50 \CIV~absorption systems separately from 115 DESI SV quasars in the COSMOS+HSC fields. All of the systems having Ly$\alpha$ within the detection limit are with log $N$(H~{\sc i})/cm$^{2}$ $>$ 14.0. In the redshift coverage of both \MGII~and \CIV~($z$ = 1.3 -- 2.5), 20 out of 34 (58.82\%) \MGII~systems have \CIV~
detection. 

2. Fourteen quasars are covered by the COSMOS2020 catalog. By cross-matching the COSMOS2020 catalog, we select the \MGII~and \CIV-galaxies above the mass limit. The majority of the \MGII~galaxies and \CIV~ galaxies are classified as the main-sequence star-forming galaxies within 0.3 dex scatter. A tight correlation between the \MGII~equivalent width and the weighted average galaxy SFR is found. The \CIV-only galaxies tentatively reside in a larger impact parameter than the systems having both \MGII~and \CIV~absorption.

3. We find that the covering fraction of strong \MGII~absorbing gas selected galaxies in main sequence galaxies is two times higher than that in all the galaxy populations within 250 kpc. The strong \MGII~contributes $\sim$0.068 star formation to the global star formation at $z\sim$ 2, which is consistent with the covering fraction of the \MGII~gas. The result suggests the co-evolution of cool gas probed by strong \MGII~and the main sequence galaxies at the cosmic noon.

4. We find that the environmental effects and the galaxy density in a large-scale structure tentatively play an active role in the origin of multiphase CGM at $z = $ 1--3.0. 

Future JWST observations in the COSMOS field will also provide more information on the galaxy morphology and star formation history of host galaxies. Methods in analyzing gaseous halo and its host galaxies in this work can be tested in the Rubin era when a large quantity of photo-$z$ will be provided.

\acknowledgments
We would like to thank the anonymous referee for very sconstructive comments. We also thank Sol\`ene Chabanier and J. Xavier Prochaska for the constructive comments in the DESI internal review. We thank Clotilde Laigle, Henry J. McCracken, John Weaver, the COSMOS team, Patrick Petitjean, Luis C. Ho, Hsiao-Wen Chen, Yunjing Wu, Ting-Wen Lan, Abhijeet Anand, Antonella Palmese and all the colleagues at the WMAG22 conference for fruitful discussions. 

SZ, LJ, and ZP acknowledge support from the National Science Foundation of China (11721303, 11890693). SZ, ZC, and ZS are supported by the National Key R\&D Program of China (grant no.\ 2018YFA0404503), the National Science Foundation of China (grant no.\ 12073014). The science research grants from the China Manned Space Project with No. CMS-CSST2021-A05, and Tsinghua University Initiative Scientific Research Program (No. 20223080023). HZ acknowledges support from the National Science Foundation of China (grant no.\ 12120101003).

This research is supported by the Director, Office of Science, Office of High Energy Physics of the U.S. Department of Energy under Contract No. DE–AC02–05CH11231, and by the National Energy Research Scientific Computing Center, a DOE Office of Science User Facility under the same contract; additional support for DESI is provided by the U.S. National Science Foundation, Division of Astronomical Sciences under Contract No. AST-0950945 to the NSF’s National Optical-Infrared Astronomy Research Laboratory; the Science and Technologies Facilities Council of the United Kingdom; the Gordon and Betty Moore Foundation; the Heising-Simons Foundation; the French Alternative Energies and Atomic Energy Commission (CEA); the National Council of Science and Technology of Mexico (CONACYT); the Ministry of Science and Innovation of Spain (MICINN), and by the DESI Member Institutions: https://www.desi.lbl.gov/collaborating-institutions.

The authors are honored to be permitted to conduct scientific research on Iolkam Du’ag (Kitt Peak), a mountain with particular significance to the Tohono O’odham Nation.

%\end{acknowledgments}

\facilities{Mayall (DESI)}
\newpage
\bibliographystyle{aasjournal}
\bibliography{./ms_final}



%\bibliographystyle{mnras}


%%%%%%%%%%%%%%%%%%%%%%%%%%%%%%%%%%%%%%%%%%%%%%%%%%

%%%%%%%%%%%%%%%%% APPENDICES %%%%%%%%%%%%%%%%%%%%%
\clearpage
\newpage
\appendix
%\section{Appendix}
\begin{figure*}[!ht]
\centering
\begin{minipage}[b]{.5\linewidth}
\centering
\includegraphics[width=\linewidth]{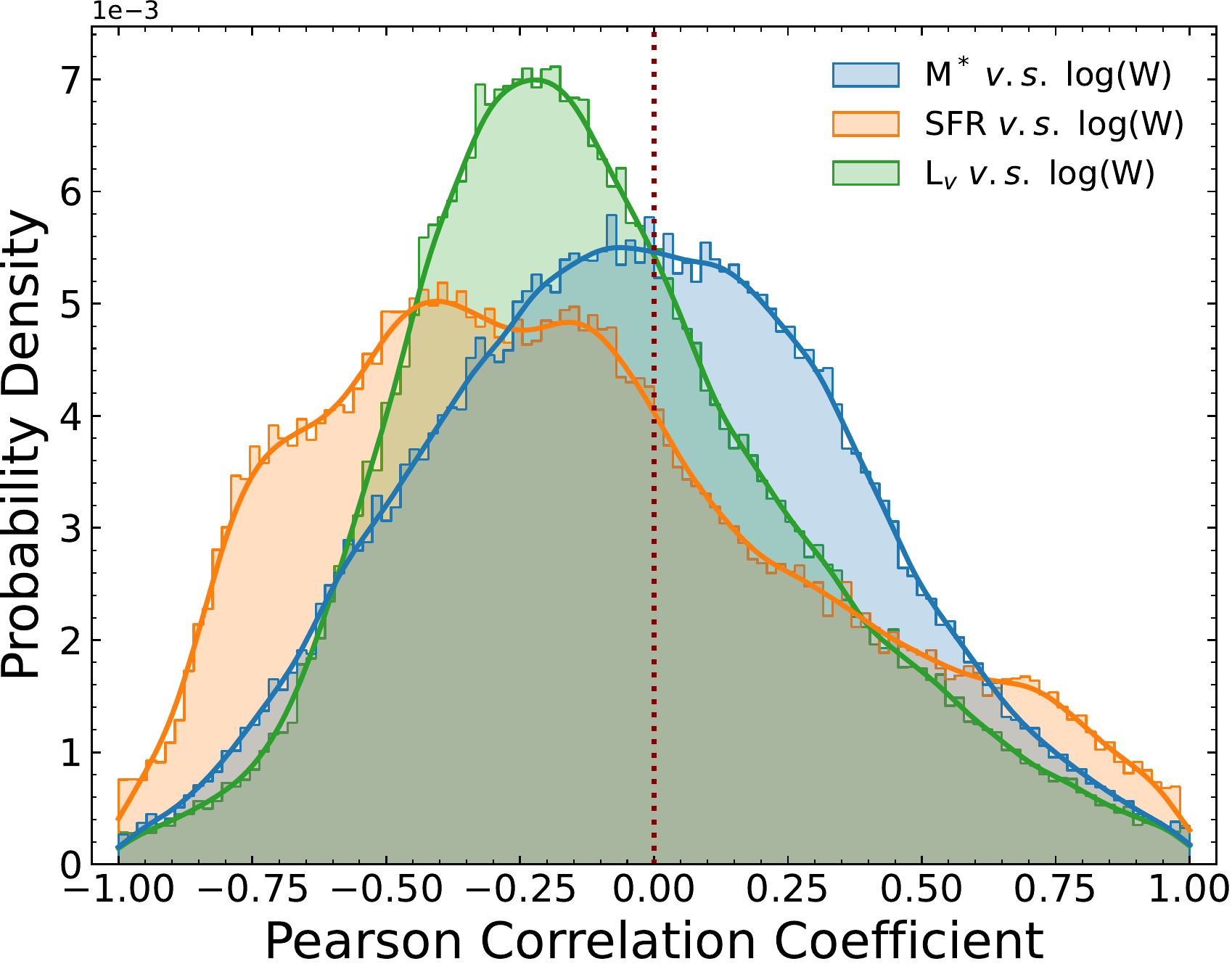}
\end{minipage}\hfill
\begin{minipage}[b]{.5\linewidth}
\centering
\includegraphics[width=\linewidth]{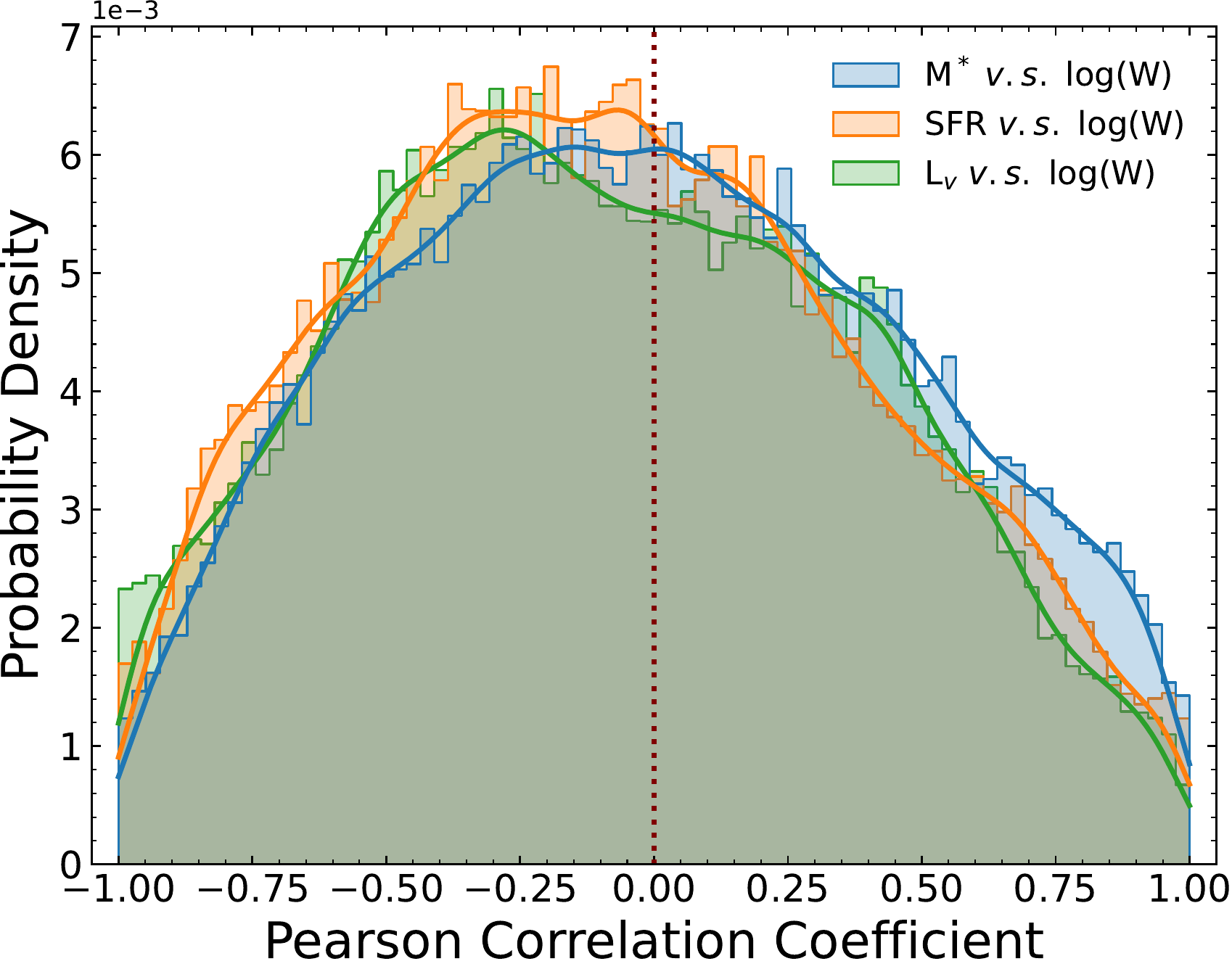}
\end{minipage}
\caption{The posterior distribution $P(r|X$) of the Pearson correlation coefficient $r$ between \MGII~(upper) and \CIV~(lower) absorber strenth with galaxy properties: log $W_r$--$M_*$ (blue, --0.03$^{+0.40}_{-0.41}$, --0.03$^{+0.55}_{-0.52}$), log $W_r$--log SFR (orange, --0.20$^{+0.57}_{-0.43}$, --0.10$^{+0.55}_{-0.50}$), and log $W_r$--$L_v$ (green, --0.14$^{+0.42}_{-0.30}$, --0.10$^{+0.55}_{-0.48}$) relations, where $X$ is the galaxy dataset. We perform a Monte Carlo simulation to sample the dataset and generate the probability density distribution.\label{fig:pearson}}
\end{figure*}

\begin{figure*}[htb!]
\gridline{
\fig{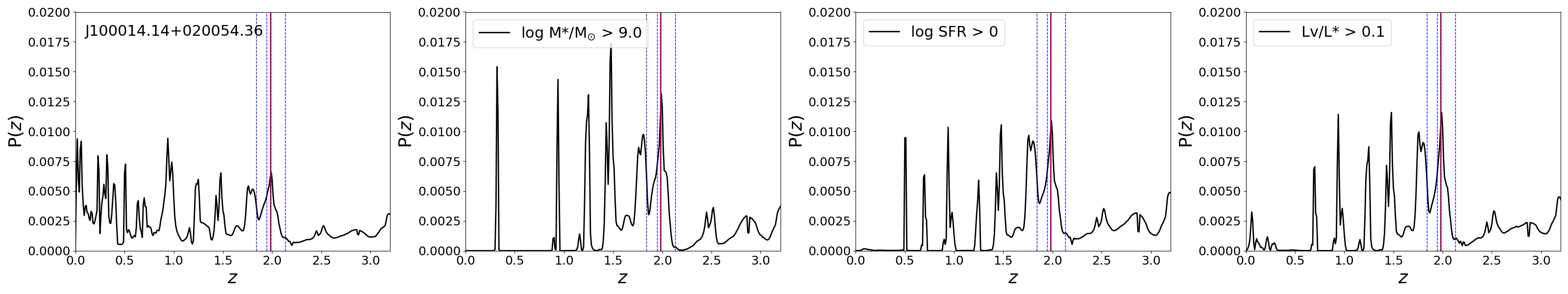}{1.0\textwidth}{}
}
\gridline{
\fig{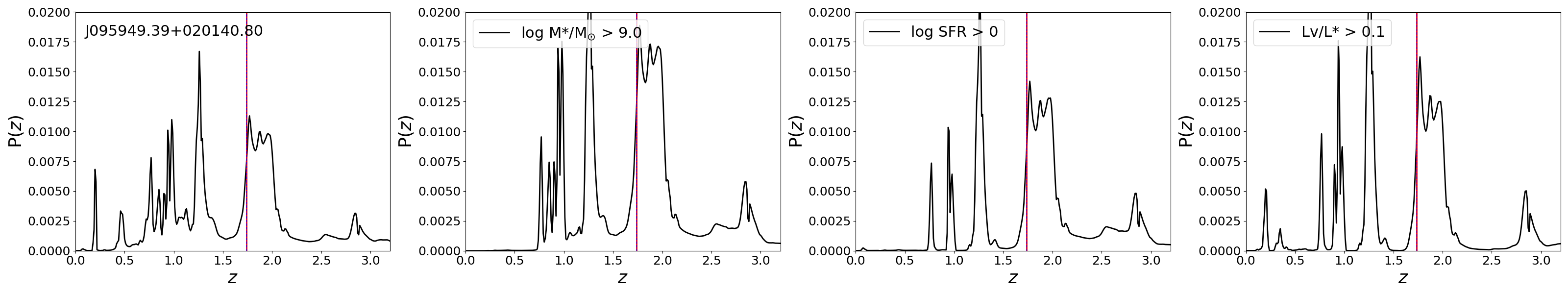}{1.0\textwidth}{}
}
\gridline{
\fig{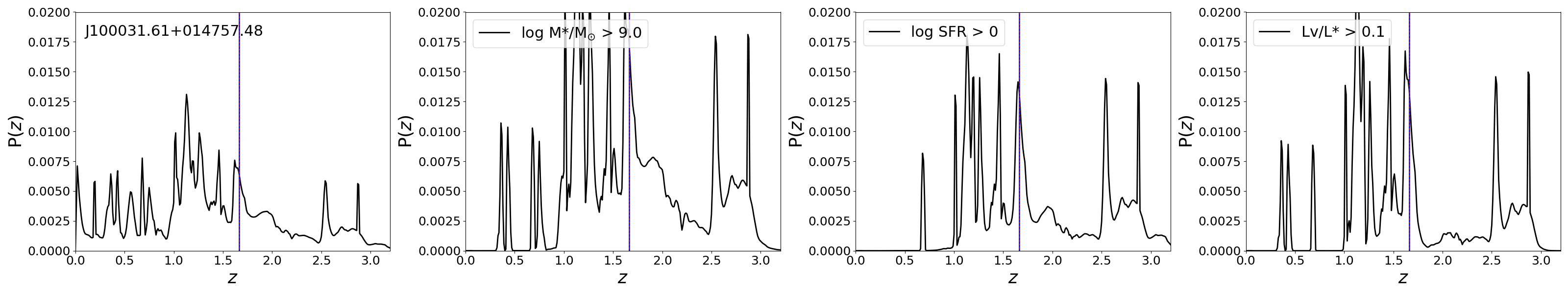}{1.0\textwidth}{}
}
\gridline{
\fig{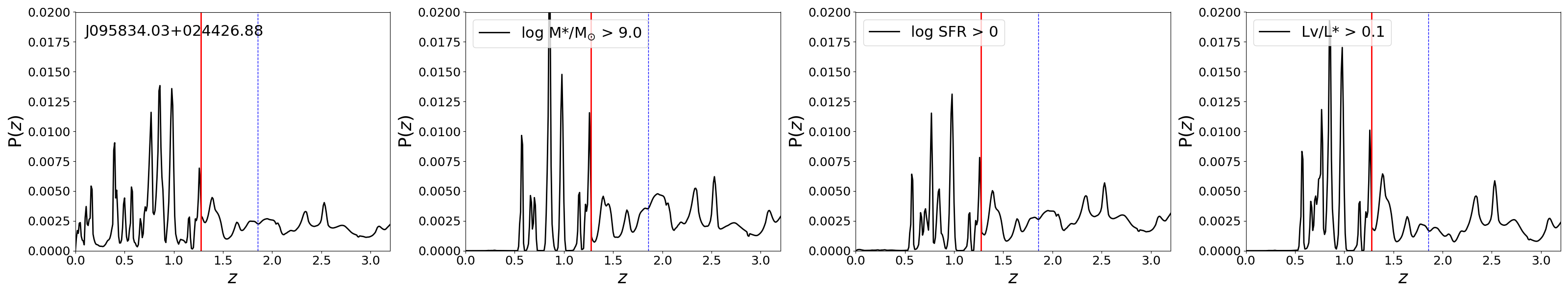}{1.0\textwidth}{}
}
\gridline{
\fig{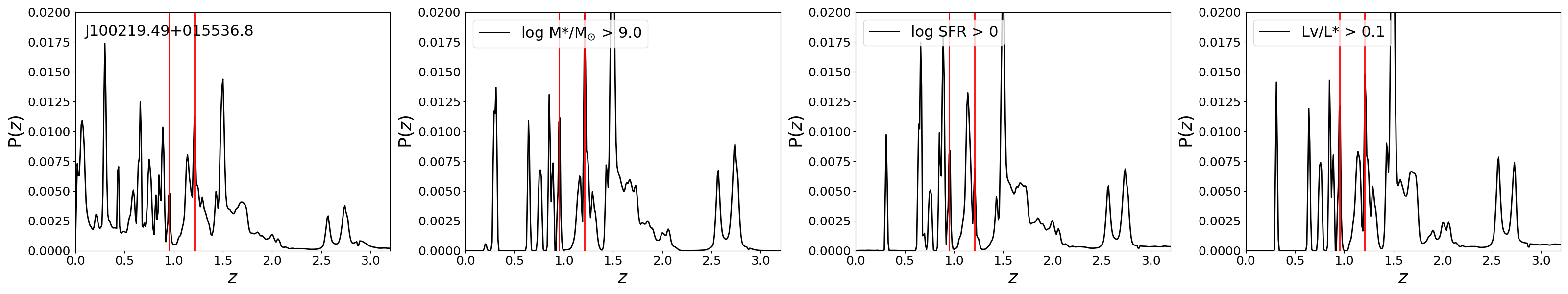}{1.0\textwidth}{}
}
\gridline{
\fig{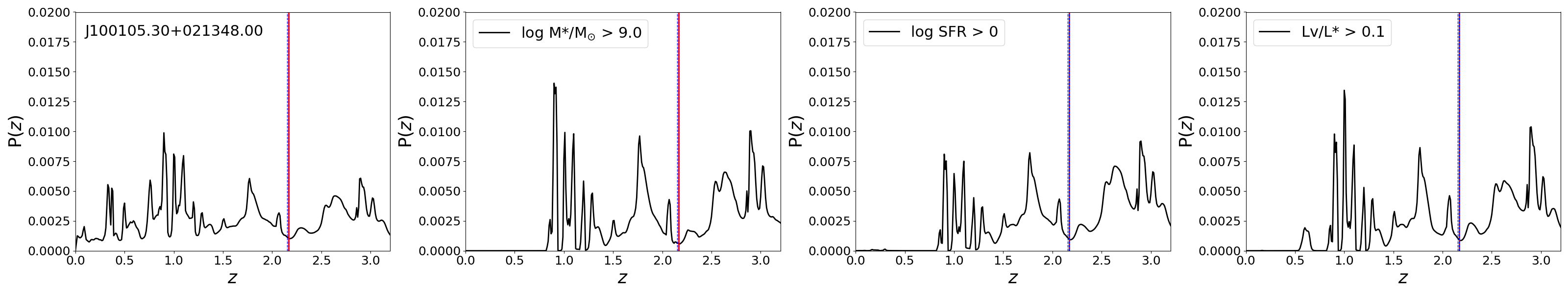}{1.0\textwidth}{}
}
\end{figure*}

\begin{figure*}
\gridline{
\fig{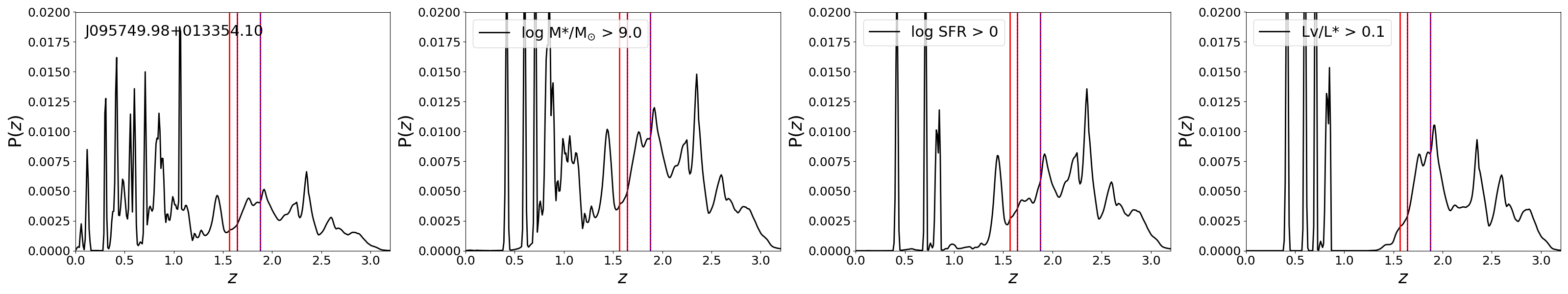}{1.0\textwidth}{}
}
\gridline{
\fig{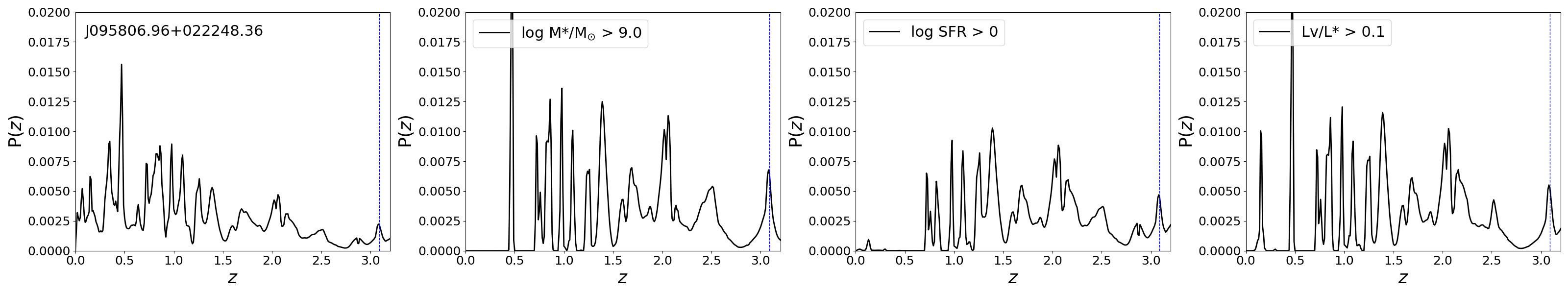}{1.0\textwidth}{}
}
\gridline{
\fig{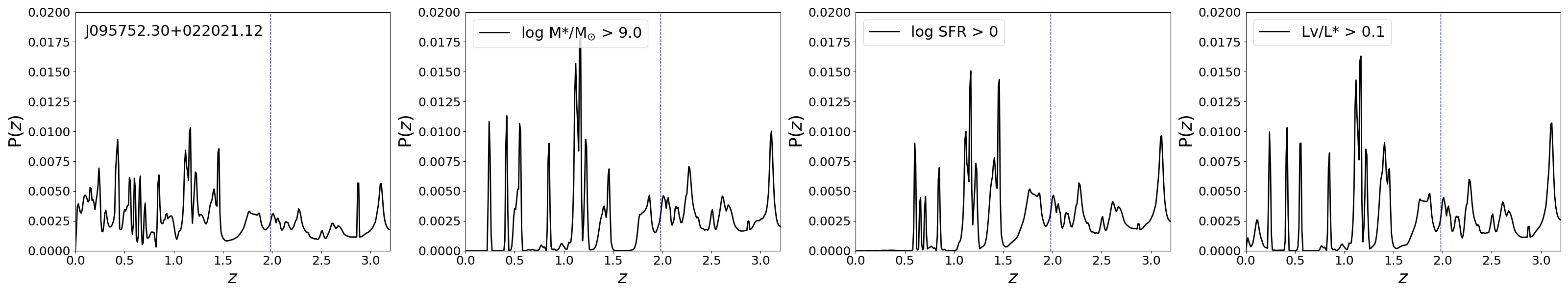}{1.0\textwidth}{}
}
\gridline{
\fig{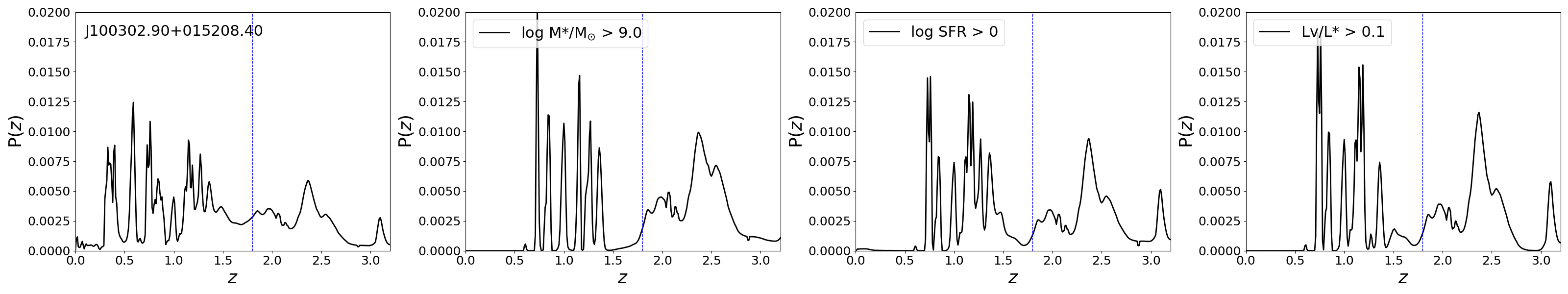}{1.0\textwidth}{}
}
\caption{Galaxy probability density P($z$) as a function of redshift. The P($z$) is a normalized photometric redshift probability distribution by taking around all the galaxies in the COSMOS2020 catalog within 30$\arcsec$ offset the quasar sightline. The red and blue dashed lines are the absorber redshifts of \MGII~and \CIV~systems. }\label{fig:g_density}
\end{figure*}

\begin{table*}
\centering
  \caption{Incidence rate $dN/dz$ and comoving density of \MGII~and \CIV~absorbers in this work. \label{table:dndz} }
\begin{tabular}{cccccc}
\hline
$\Delta z$  & \MGII~ $dN/dz$  &  \MGII~ $dN/dz$&   $\Delta z$  & \CIV~ $dN/dz$ &\CIV~$dN/dz$ \\
    &($W_r>$ 0.3 \AA) &($W_r>$ 1.0 \AA) & & ($W_r>$ 0.2 \AA) &($W_r>$ 1.0 \AA) \\
\hline
 0.60 -- 1.00   & 0.606 $\pm$ 0.076 &   0.275 $\pm$ 0.127 & 1.30 -- 1.50 &  1.675 $\pm$ 0.149  &   0.494 $\pm$ 0.094    \\ 
 1.00 -- 1.50   & 1.079 $\pm$ 0.119 &   0.484 $\pm$ 0.293 & 1.50 -- 2.00 &  1.549 $\pm$ 0.206  &   0.420 $\pm$ 0.122    \\   1.50 -- 2.00   & 1.330 $\pm$ 0.187 & 0.602 $\pm$ 0.141 & 2.00 -- 2.50 &  2.311 $\pm$ 0.447  &  0.800 $\pm$ 0.298    \\ 
 2.00 -- 2.50   & 1.440 $\pm$ 0.298 &   0.403 $\pm$ 0.149 & 2.50 -- 3.20 &  3.689 $\pm$ 0.1.07 & 1.618 $\pm$ 0.799    \\
\hline
\end{tabular}
\end{table*}%

\startlongtable
\centering
\begin{deluxetable*}{llllll}
\tablecaption{Quasars used in this work. (1) Quasar name (2) Quasar emission redshift (3) RA (4) DEC (5) The quasar effective exposure time (6) The mean signal-to-noise measured from three continuum region without obvious emission or absorption.}\label{table:qso_info}
\tablehead{\colhead{TARGETID(1)} & \colhead{$z_{em}$(2)} & \colhead{RA(3)} & \colhead{DEC(4)}  & \colhead{EXPTIME(5)} & \colhead{S/N(6)}}
\startdata
J095426.83+025022.92 &  1.3422   &  148.6118 &  2.0897 &  22884.44 &  8   \\  
J095430.38+021525.92 &  2.0498   &  148.6266 &  2.2572 &  7302.19  &  17  \\  
J095435.15+023142.24 &  1.3138   &  148.6465 &  2.5284 &  22884.44 &  7   \\  
J095436.38+027031.08 &  1.8071   &  148.6516 &  2.1253 &  22884.44 &  7   \\  
J095446.03+014639.00 &  1.1153   &  148.6918 &  1.7775 &  7302.19  &  3   \\  
J095458.24+015616.79 &  0.7464   &  148.7427 &  1.9380 &  22884.44 &  29  \\  
J095504.24+026052.92 &  3.0927   &  148.7677 &  2.1147 &  22884.44 &  6   \\  
J095505.44+021028.92 &  1.2235   &  148.7727 &  2.1747 &  22884.44 &  4   \\  
J095525.79+028011.76 &  3.1414   &  148.8575 &  2.1366 &  2884.447 &  3   \\  
J095615.91+024652.67 &  1.6542   &  149.0663 &  2.7813 &  22884.44 &  14  \\  
J095656.18+021314.88 &  1.1215   &  149.2341 &  2.2208 &  22884.44 &  37  \\  
J095712.88+014917.39 &  1.1818   &  149.3037 &  1.8215 &  22884.44 &  12  \\  
J095726.32+024027.83 &  0.9583   &  149.3597 &  2.0744 &  22884.44 &  22  \\  
J095739.24+015533.23 &  1.8146   &  149.4135 &  1.9259 &  22884.44 &  3   \\   
J095749.96+013353.99 &  2.0055   &  149.4582 &  1.5650 &  7302.19  &  18  \\   
J095752.29+022021.11 &  2.0490   &  149.4679 &  2.3392 &  7302.19  &  41  \\   
J095806.96+022248.36 &  3.0956   &  149.5290 &  2.3801 &  7302.19  &  6   \\   
J095820.44+023003.95 &  1.3578   &  149.5852 &  2.0511 &  22884.44 &  36  \\   
J095834.03+024426.88 &  1.8927   &  149.6418 &  2.7408 &  7302.19  &  33  \\   
J095834.51+034338.63 &  1.2651   &  149.6438 &  3.7274 &  7302.19  &  7   \\   
J095847.11+035003.84 &  1.8582   &  149.6963 &  3.8344 &  22884.44 &  15  \\   
J095900.11+033651.83 &  1.5873   &  149.7505 &  3.6144 &  7302.19  &  4   \\   
J095910.94+019046.80 &  2.6908   &  149.7956 &  1.1630 &  22884.44 &  5   \\   
J095911.63+033442.95 &  1.8081   &  149.7985 &  3.5786 &  22884.44 &  3   \\   
J095915.28+034033.23 &  0.6818   &  149.8137 &  3.6759 &  7302.19  &  19  \\   
J095922.27+034046.56 &  4.0667   &  149.8428 &  3.6796 &  7302.19  &  3   \\   
J095923.80+003853.16 &  0.7838   &  149.8492 &  0.6481 &  7302.19  &  3   \\   
J095931.72+033710.19 &  1.1300   &  149.8822 &  3.6195 &  6402.13  &  4   \\   
J095933.12+033118.12 &  2.1464   &  149.8880 &  3.5217 &  7302.19  &  6   \\   
J095946.82+004918.47 &  2.2488   &  149.9451 &  0.8218 &  22884.44 &  70  \\   
J095949.39+021040.80 &  1.7533   &  149.9558 &  2.0280 &  7302.19  &  26  \\   
J095956.56+004301.55 &  1.9471   &  149.9857 &  0.7171 &  7302.19  &  4   \\   
J100009.35+005311.03 &  0.9103   &  150.0390 &  0.8864 &  22884.44 &  6   \\   
J100011.59+004154.24 &  1.3761   &  150.0483 &  0.6984 &  7302.19  &  3   \\   
J100014.13+020054.35 &  2.4968   &  150.0589 &  2.0151 &  22884.44 &  32  \\   
J100017.87+005400.00 &  0.7289   &  150.0745 &  0.9000 &  7302.19  &  5   \\   
J100020.49+015011.40 &  1.5242   &  150.0854 &  1.0865 &  22884.44 &  8   \\   
J100020.49+033247.76 &  2.0233   &  150.0854 &  3.5466 &  22884.44 &  3   \\   
J100022.72+033724.96 &  0.9057   &  150.0947 &  3.6236 &  7302.19  &  6   \\   
J100023.78+035500.12 &  1.1259   &  150.0991 &  3.9167 &  22884.44 &  16  \\   
J100025.32+034823.76 &  2.4103   &  150.1055 &  3.8066 &  22884.44 &  15  \\   
J100029.13+011044.75 &  1.0591   &  150.1214 &  1.0291 &  7302.19  &  37  \\   
J100029.68+035023.27 &  1.6433   &  150.1237 &  3.0898 &  22884.44 &  21  \\   
J100031.60+014757.47 &  1.683    &  150.1317 &  1.7993 &  22884.44 &  17  \\   
J100032.80+033458.43 &  1.724    &  150.1367 &  3.5829 &  7302.19  &  6   \\   
J100036.98+018003.12 &  1.8355   &  150.1541 &  1.1342 &  7302.19  &  3   \\   
J100037.39+034455.68 &  3.2336   &  150.1558 &  3.7488 &  7302.19  &  15  \\   
J100038.47+015009.24 &  2.1797   &  150.1603 &  1.0859 &  22884.44 &  9   \\   
J100039.23+012006.35 &  3.8343   &  150.1635 &  1.0351 &  7302.19  &  5   \\   
J100039.55+033216.44 &  0.9768   &  150.1648 &  3.5379 &  22884.44 &  11  \\ 
J100042.11+034911.27 &  1.6542   &  150.1755 &  3.8198 &  22884.44 &  12  \\ 
J100042.98+004438.76 &  1.452    &  150.1791 &  0.7441 &  7302.19  &  5   \\ 
J100043.39+033217.16 &  2.2667   &  150.1808 &  3.5381 &  22884.44 &  3   \\ 
J100047.92+034700.95 &  1.9006   &  150.1997 &  3.7836 &  7302.19  &  3   \\ 
J100048.83+005925.79 &  0.8704   &  150.2035 &  0.9905 &  7302.19  &  3   \\ 
J100048.83+033039.23 &  3.3652   &  150.2035 &  3.5109 &  22884.44 &  7   \\ 
J100052.29+005121.59 &  1.9461   &  150.2179 &  0.8560 &  7302.19  &  7   \\ 
J100053.80+033105.87 &  2.4587   &  150.2242 &  3.5183 &  22884.44 &  3   \\ 
J100055.00+005508.40 &  2.0652   &  150.2292 &  0.9190 &  7302.19  &  5   \\ 
J100058.53+004837.44 &  1.7392   &  150.2439 &  0.8104 &  7302.19  &  3   \\ 
J100058.82+015359.99 &  1.5619   &  150.2451 &  1.9000 &  7302.19  &  16  \\ 
J100101.03+035233.96 &  2.7714   &  150.2543 &  3.8761 &  22884.44 &  1   \\ 
J100104.60+004648.00 &  2.5726   &  150.2692 &  0.7800 &  7302.19  &  3   \\ 
J100105.30+021347.99 &  2.6066   &  150.2721 &  2.2300 &  7302.19  &  8   \\ 
J100106.43+033650.39 &  2.0931   &  150.2768 &  3.6140 &  7302.19  &  4   \\ 
J100108.66+005730.59 &  2.0411   &  150.2861 &  0.9585 &  7302.19  &  25  \\ 
J100109.21+004859.40 &  0.6574   &  150.2884 &  0.8165 &  7302.19  &  8   \\ 
J100111.40+034100.23 &  2.2668   &  150.2975 &  3.6834 &  7302.19  &  29  \\ 
J100111.49+033506.36 &  2.4574   &  150.2979 &  3.5851 &  7302.19  &  21  \\ 
J100113.39+005009.95 &  2.5853   &  150.3058 &  0.8361 &  7302.19  &  4   \\ 
J100118.31+035101.07 &  2.8947   &  150.3263 &  3.8503 &  22884.44 &  6   \\ 
J100125.46+005205.15 &  0.7802   &  150.3561 &  0.8681 &  22884.44 &  56  \\ 
J100126.59+004648.71 &  1.7273   &  150.3608 &  0.7802 &  7302.19  &  8   \\ 
J100127.55+034434.07 &  2.8133   &  150.3648 &  3.7428 &  22884.44 &  22  \\ 
J100129.44+003813.55 &  2.9074   &  150.3727 &  0.6371 &  4602.01  &  3   \\ 
J100132.08+005259.15 &  1.0669   &  150.3837 &  0.8831 &  22884.44 &  8   \\ 
J100133.36+005118.71 &  1.403    &  150.3890 &  0.8552 &  22884.44 &  11  \\ 
J100134.15+011021.72 &  1.7589   &  150.3923 &  1.1727 &  7302.19  &  3   \\
J100136.36+034309.48 &  1.1248   &  150.4015 &  3.7193 &  5502.06  &  5   \\
J100137.19+021612.35 &  1.65     &  150.4050 &  2.2701 &  2884.447 &  11  \\
J100137.77+004655.56 &  2.5841   &  150.4074 &  0.7821 &  7302.19  &  25  \\
J100138.97+033616.19 &  1.2473   &  150.4124 &  3.6045 &  7302.19  &  10  \\
J100140.31+003947.52 &  1.6027   &  150.4180 &  0.6632 &  7302.19  &  4   \\
J100142.04+003907.55 &  1.3516   &  150.4252 &  0.6521 &  7302.19  &  3   \\
J100142.55+015031.20 &  1.8225   &  150.4273 &  1.0920 &  7302.19  &  6   \\
J100147.88+021447.03 &  0.8804   &  150.4495 &  2.2464 &  22884.44 &  17  \\
J100205.23+004249.68 &  1.7855   &  150.5218 &  0.7138 &  7302.19  &  11  \\
J100217.87+004252.20 &  1.2343   &  150.5745 &  0.7145 &  7302.19  &  10  \\
J100219.48+015536.84 &  1.5087   &  150.5812 &  1.9269 &  7302.19  &  17  \\
J100236.69+015948.47 &  1.5192   &  150.6529 &  1.9968 &  7302.19  &  19  \\
J100302.90+015208.40 &  1.8026   &  150.7621 &  1.8690 &  7302.19  &  22  \\
J100344.35+025002.03 &  2.9914   &  150.9348 &  2.0839 &  7302.19  &  8   \\
J100348.67+021044.76 &  1.3908   &  150.9528 &  2.1791 &  7302.19  &  4   \\
J100417.61+021330.35 &  3.1056   &  151.0734 &  2.2251 &  22884.44 &  3   \\
J100441.78+021147.04 &  1.828    &  151.1741 &  2.1964 &  22884.44 &  6   \\
J100449.99+021641.52 &  1.7882   &  151.2083 &  2.2782 &  22884.44 &  3   \\
J100505.03+021519.08 &  2.8624   &  151.2710 &  2.2553 &  22884.44 &  9   \\
J100520.87+021112.84 &  2.3843   &  151.3370 &  2.1869 &  7302.19  &  7   \\
J100523.85+015920.40 &  1.7769   &  151.3494 &  1.9890 &  22884.44 &  18  \\
J100524.86+025047.76 &  1.0844   &  151.3536 &  2.0966 &  22884.44 &  6   \\
J100527.09+027025.31 &  1.5706   &  151.3629 &  2.1237 &  22884.44 &  5   \\
J100534.43+021015.96 &  1.8209   &  151.3935 &  2.1711 &  7302.19  &  18  \\
J100541.51+015950.64 &  1.7298   &  151.4230 &  1.9974 &  22884.44 &  16  \\
J100542.69+021516.92 &  1.1336   &  151.4279 &  2.2547 &  22884.44 &  4   \\
J100546.20+027052.67 &  1.7086   &  151.4425 &  2.1313 &  22884.44 &  4   \\
J100547.68+021221.59 &  0.9066   &  151.4487 &  2.2060 &  22884.44 &  11  \\
J100606.45+021445.23 &  1.1864   &  151.5269 &  2.2459 &  22884.44 &  6   \\
J100624.45+014758.20 &  1.0138   &  151.6019 &  1.7995 &  22884.44 &  4   \\
J100632.71+013955.43 &  2.4867   &  151.6363 &  1.6654 &  6402.13  &  3   \\
J100634.60+026014.40 &  2.3479   &  151.6442 &  2.1040 &  7302.19  &  3   \\
J100638.88+014941.16 &  2.1134   &  151.6620 &  1.8281 &  7302.19  &  3   \\
J100641.11+021658.07 &  2.0098   &  151.6713 &  2.2828 &  22884.44 &  17  \\
J100039.25+010206.47 &  3.8344   &  150.1635 &  1.0350 &  7302.19  &  18 \\
J100015.90+010801.75 &  2.0098   &  151.6713 & 2.2828 & 22884.44   & 17   \\
\enddata
\end{deluxetable*}

\clearpage
\newpage

\startlongtable
\centering
\begin{deluxetable*}{llllll}
\tablecaption{Metal absorbers Measurements (1) Quasar name (2) Absorber redshift (3) Velocity width of \MGII~($\lambda$2796) line (4) Equivalent width of the \MGII~($\lambda$2796) line (5) Velocity width of \CIV~($\lambda$1548) line (6) Equivalent width of the \CIV~($\lambda$1548) line.}\label{table:metal_measurement}
\tablehead{\colhead{TARGETID(1)} & \colhead{$z_{abs}$(2)} & \colhead{$\Delta v(\lambda2796)$(3)} & \colhead{ $W_r(\lambda2796)$(4)}  & \colhead{$\Delta (\lambda1548)$(5)} &  \colhead{$W_r(\lambda1548)$(6)}\\
\colhead{} & \colhead{} & \colhead{(km s$^{-1}$)} & \colhead{(\AA)}  & \colhead{(km s$^{-1}$}) &  \colhead{(\AA)}}
\startdata
J100219.48+015536.84 & 0.9520   &  495.462  &  1.462$\pm$0.316   &   nan      & nan             \\   
J100219.48+015536.84 & 1.2090   &  206.951  &  0.631$\pm$0.096   &   nan      & nan             \\   
J100302.90+015208.40 & 1.7970   &  411.110  &  0.533$\pm$0.091   &   nan      & nan             \\
J095749.96+013353.99 & 1.6428   &  379.000  &  1.223$\pm$0.162   &   419.351  & 0.738$\pm$0.2     \\
J095749.96+013353.99 & 1.5660   &  465.260  &  1.169$\pm$0.202   &   375.754  & 0.716$\pm$0.142   \\
J095749.96+013353.99 & 1.8770   &  314.938  &  1.234$\pm$0.114   &   nan      & nan             \\      
J095834.03+024426.88 & 1.2756   &  393.437  &  1.005$\pm$0.126   &   nan      & nan             \\      
J095820.44+023003.95 & 0.7505   &  480.986  &  0.9  $\pm$0.087   &   nan      & nan             \\      
J095808.15+015425.20 & 2.4845   &  119.614  &  1.676$\pm$0.666   &   nan      & nan             \\      
J100113.39+005009.95 & 1.0620   &  294.506  &  1.853$\pm$0.636   &   nan      & nan             \\      
J100137.77+004655.56 & 2.1385   &  528.224  &  1.656$\pm$0.278   &   nan      & nan             \\      
J100137.77+004655.56 & 1.3445   &  235.737  &  0.41 $\pm$0.108   &   nan      & nan             \\      
J100205.23+004249.68 & 1.2510   &  410.364  &  1.286$\pm$0.392   &   nan      & nan             \\      
J100039.25+010206.47 & 1.3450   &  219.637  &  0.74 $\pm$0.385   &   nan      & nan             \\      
J100108.66+005730.59 & 2.0250   &  nan      &  nan               &   402.453  & 0.386$\pm$0.065   \\
J100108.66+005730.59 & 1.7378   &  486.468  &  0.570$\pm$0.196   &   449.654  & 0.503$\pm$0.193   \\
J095430.38+021525.92 & 1.8315   &  636.834  &  0.636$\pm$0.43    &   364.219  & 0.84 $\pm$0.189   \\
J095430.38+021525.92 & 1.1072   &  336.301  &  0.346$\pm$0.22    &   nan      & nan             \\              
J100111.40+034100.23 & 2.2660   &  nan      &  nan               &   606.366  & 1.114$\pm$0.055   \\
J100111.40+034100.23 & 1.4660   &  nan      &  nan               &   976.981  & 0.625$\pm$0.408   \\
J100111.40+034100.23 & 1.5475   &  431.816  &  0.476$\pm$0.157   &   nan      & nan             \\       
J100111.40+034100.23 & 1.6715   &  352.731  &  0.336$\pm$0.139   &   nan      & nan             \\       
J100534.43+021015.96 & 1.4990   &  693.396  &  4.891$\pm$0.283   &   nan      & nan             \\       
J100534.43+021015.96 & 1.1940   &  391.394  &  0.719$\pm$0.195   &   nan      & nan             \\       
J100111.49+033506.36 & 1.4140   &  nan      & nan                &   643.073  & 0.504$\pm$0.217    \\
J100111.49+033506.36 & 0.6605   &  785.408  &  0.728$\pm$0.36    &   nan      & nan             \\       
J100111.49+033506.36 & 1.4140   &  444.960  &  0.712$\pm$0.16    &   nan      & nan             \\       
J100541.51+015950.64 & 0.8280   &  420.354  &  0.908$\pm$0.278   &   nan      & nan             \\       
J100541.51+015950.64 & 1.4480   &  316.473  &  1.408$\pm$0.155   &   451.940  & 1.444$\pm$0.268   \\
J100527.09+027025.31 & 1.4846   &  360.323  &  1.699$\pm$0.574   &   464.818  & 1.591$\pm$0.559   \\
J100042.11+034911.27 & 1.6425   &  544.964  &  2.611$\pm$0.141   &   610.488  & 1.907$\pm$0.261   \\
J095946.82+004918.47 & 2.1544   &  359.402  &  0.399$\pm$0.034   &   482.318  & 0.227$\pm$0.043   \\
J095847.11+035003.84 & 1.3310   &  453.723  &  0.856$\pm$0.203   &   nan      & nan             \\       
J100025.32+034823.76 & 1.5140   &  349.751  &  0.741$\pm$0.205   &   374.901  & 0.627$\pm$0.079   \\
J100029.68+035023.27 & 0.9760   &  423.160  &  1.01 $\pm$0.196   &   nan      & nan             \\        
J100217.87+004252.20 & 1.2230   &  213.961  &  0.274$\pm$0.149   &   nan      & nan             \\        
J100015.90+010801.75 & 1.4780   &  312.211  &  2.061$\pm$0.8     &   nan      & nan             \\        
J100632.71+013955.43 & 2.4886   &  146.668  &  0.848$\pm$0.249   &   nan      & nan             \\        
J095949.39+021040.80 & 1.7372   &  530.321  &  1.152$\pm$0.175   &   522.252  & 1.126$\pm$0.08    \\
J100058.82+015359.99 & 0.6715   &  603.396  &  2.235$\pm$0.36    &   nan      & nan             \\              
J100105.30+021347.99 & 2.1680   &  nan      &  nan               &   585.025  &  1.102$\pm$0.421   \\
J100105.30+021347.99 & 2.1530   &  212.247  &  0.799$\pm$0.212   &   478.487  & 0.518$\pm$0.341   \\
J100014.13+020054.35 & 1.4700   &  241.531  &  0.804$\pm$0.332   &   nan      & nan             \\       
J100133.36+005118.71 & 0.7838   &  580.924  &  1.552$\pm$0.531   &   nan      & nan             \\       
J100031.60+014757.47 & 1.6625   &  689.316  &  1.836$\pm$0.257   &   722.683  & 1.828$\pm$0.27    \\
J095426.83+025022.92 & 1.3500   &  403.319  &  1.279$\pm$0.253   &   504.614  & 0.877$\pm$0.281   \\
J095726.32+024027.83 & 0.7575   &  381.720  &  0.622$\pm$0.158   &   nan      & nan             \\             
J100014.13+020054.35 & 2.1305   &  nan      &  nan               &   395.208  & 0.434$\pm$0.071   \\
J100014.13+020054.35 & 1.9813   &  342.842  &  0.300$\pm$0.117   &   514.078  & 0.453$\pm$0.148   \\
J100014.13+020054.35 & 1.9450   &  nan      &  nan               &   558.430  & 0.62 $\pm$0.099   \\
J100014.13+020054.35 & 1.8400   &  nan      &  nan               &   568.266  & 0.574$\pm$0.147   \\
J100137.19+021612.35 & 1.6360   &  497.133  &  2.682$\pm$0.205   &   669.046  & 2.911$\pm$0.99    \\
J100441.78+021147.04 & 1.4015   &  416.102  &  1.794$\pm$0.521   &   nan      & nan             \\                 
J095435.15+023142.24 & 1.3010   &  443.636  &  1.971$\pm$0.336   &   nan      & nan             \\                    
J100127.55+034434.07 & 2.7740   &  nan      &  nan               &   422.899  & 1.34$\pm$0.09    \\       
J100127.55+034434.07 & 2.2294   &  284.891  &  0.327$\pm$0.122   &   468.897  & 1.08$\pm$0.125   \\
J100055.00+005508.40 & 1.9490   &  447.044  &  1.538$\pm$0.517   &   nan      & nan             \\              
J095752.29+022021.11 & 1.9810   &  nan      &  nan               &   523.836  & 0.546$\pm$0.210   \\
J095834.03+024426.88 & 1.8555   &  nan      &  nan               &   508.977  & 0.252$\pm$0.103   \\
J095826.64+024228.00 & 2.4520   &  nan      &  nan               &   482.868  & 0.923$\pm$0.296   \\
J095839.84+024424.00 & 3.1620   &  nan      &  nan               &   220.482  & 0.530$\pm$0.100   \\
J095839.84+024424.00 & 3.1800   &  nan      &  nan               &   152.901  & 0.217$\pm$0.100   \\
J100038.47+015009.24 & 2.1774   &  nan      &  nan               &   510.516  & 1.124$\pm$0.121   \\
J100104.60+004648.00 & 2.5350   &  nan      &  nan               &   602.860  & 3.052$\pm$0.991   \\
J100129.44+003813.55 & 2.9100   &  nan      &  nan               &   515.093  & 1.92 $\pm$0.482   \\
J100055.00+005508.40 & 2.0710   &  nan      &  nan               &   464.685  & 1.506$\pm$0.271   \\
J100638.88+014941.16 & 2.1080   &  nan      &  nan               &   323.381  & 0.922$\pm$0.214   \\
J100344.35+025002.03 & 3.0290   &  nan      &  nan               &   481.671 & 1.477$\pm$0.258   \\
J095806.96+022248.36 & 3.0880   &  nan      &  nan               &   369.762  & 1.207$\pm$0.211   \\
J095933.12+033118.12 & 2.1020   &  nan      &  nan               &   504.720  & 1.098$\pm$0.35    \\
J095933.12+033118.12 & 2.1320   &  248.939  &  0.859$\pm$0.217   &   581.047  & 1.143$\pm$0.202   \\
J100037.39+034455.68 & 2.8240   &  nan      &  nan               &   400.547  & 0.56 $\pm$0.123   \\
J095504.24+026052.92 & 2.2050   &  nan      &  nan               &   347.999  & 0.694$\pm$0.103   \\
J095436.38+027031.08 & 1.7950   &  nan      &  nan               &   721.820  & 2.092$\pm$0.246   \\
J095525.79+028011.76 & 3.1390   &  nan      &  nan               &   261.151  & 0.684$\pm$0.133   \\
J100449.99+021641.52 & 1.7362   &  nan      &  nan               &   457.749  & 1.303$\pm$0.137   \\
J100020.49+033247.76 & 2.0110   &  nan      &  nan               &   360.158  & 1.18 $\pm$0.639   \\
J100043.39+033217.16 & 2.2674   &  nan      &  nan               &   480.610  & 1.871$\pm$0.278   \\
\enddata
\end{deluxetable*}

% Don't change these lines
%\bsp	% typesetting comment

%\subsection{Notes on Individual Systems}

%{\bf 39627871785848171}

%{\bf 39627847635042937} 

%Five components 

%{\bf 39627805369044515}

%{\bf 39627877817256029} on top MgII, MgI, clear FeII

\end{document}